\pdfoutput=1
\documentclass[%
 nofootinbib,
 reprint,
 amsmath,
 amssymb,
 aps,
 prb,
 superscriptaddress,
 floatfix
]{revtex4-2}

\usepackage{graphicx}
\usepackage{dcolumn}
\usepackage{bm}
\newcommand{\ket}[1]{|{#1}\rangle}
\newcommand{\bra}[1]{\langle{#1}|}
\newcommand{\braket}[1]{\langle{#1}\rangle}
\usepackage[caption=false]{subfig}

\usepackage{mathtools}
\usepackage[T1]{fontenc}
\usepackage[utf8]{inputenc}
\usepackage{textcomp}
\usepackage[english]{babel}
\usepackage{lmodern}
\usepackage[babel]{microtype}

\DeclareUnicodeCharacter{2013}{--} % breaking at ndash

\usepackage{xcolor}
\usepackage[colorlinks,
linkcolor=black,
citecolor=black,
filecolor=black,
urlcolor=blue!50!black,
pdfusetitle]{hyperref}
\hypersetup{
pdfauthor={Axel Fünfhaus, Thilo Kopp, Elias Lettl},
pdfsubject={Solid-State Physics},
pdfkeywords={topology, vortex fields, topological defects,
  winding vectors, Hofstadter model}
 }
\newcommand*\diff{\mathop{}\!\mathrm{d}}

\let\vec\boldsymbol
\DeclareMathOperator{\Arg}{Arg}
\def\pdagger{\phantom{\dagger}}
\let\Phi\varPhi
\let\Psi\varPsi
\def\ci{\mathrm{i}}
\DeclareMathOperator{\e}{e}

\begin{document}

\title{Winding vectors of topological defects: Multiband Chern numbers}

\author{Axel Fünfhaus}
\email{fuenfhaus@itp.uni-frankfurt.de}
\affiliation{
 Institute of Theoretical Physics, Goethe University Frankfurt, Max-von-Laue-Straße 1, 60438 Frankfurt am Main, Germany
}
\affiliation{
 Center for Electronic Correlations and Magnetism, Experimental Physics VI, Institute of Physics, University of Augsburg, 86135 Augsburg, Germany
}
\author{Thilo Kopp}
\author{Elias Lettl}
\affiliation{
 Center for Electronic Correlations and Magnetism, Experimental Physics VI, Institute of Physics, University of Augsburg, 86135 Augsburg, Germany
}

\date{\today}

\begin{abstract}
Chern numbers can be calculated within a frame of vortex fields related to phase conventions of a wave function. In a band protected by gaps the Chern number is equivalent to the total number of flux carrying vortices. In the presence of topological defects like Dirac cones this method becomes problematic, in particular if they lack a well-defined winding number. We develop a scheme to include topological defects into the vortex field frame. A winding number is determined by the behavior of the phase in reciprocal space when encircling the defect's contact point. To address the possible lack of a winding number we utilize a more general concept of winding vectors. We demonstrate the usefulness of this ansatz on Dirac cones generated from bands of the Hofstadter model.
\end{abstract}

\maketitle

\section{Introduction}\label{sec:Introduction}

Chern numbers characterize topologically invariant properties of two-dimensional insulators~\cite{avr}. They are computed within the Berry formalism~\cite{sim}. In the case of gapped bands the problem reduces to computing the Berry curvature and integrating it over a suitable closed manifold such as the Brillouin zone. Instead of analyzing the Berry curvature one can also use a frame of vortex fields of the complex wave functions~\cite{koh, hat2, hat}: The manifold is divided into patches, in which the wave function has to be defined uniquely according to an appropriate phase convention. One then rewrites the integral over the Berry curvature using Stokes' theorem as a line integral along the edges of these patches. This results in determining the winding number of the phase of the transition function $\e^{\ci \chi(\vec{k})}$, which relates the phases of the wave functions in adjacent patches with phase difference $\chi(\vec{k})$. Extending $\chi(\vec{k})$ over the entire manifold generates a field of vortices, the number of which can readily be counted and is equal to the Chern number. This method has been successfully applied to calculate the Chern numbers of free electrons in a magnetic field~\cite{fra}, electrons in a periodic potential~\cite{koh, koh2, hat3} or fractional Chern insulators~\cite{haf, ger} and other systems with degenerate ground state multiplets~\cite{hat, hat4}.

Problems arise, once Bloch bands become gapless due to Dirac cones. In that case it remains to compute the Chern number for the connected bands jointly. The Berry formalism then has to be modified. One has to work with a non-Abelian Berry holonomy instead of a mere phase factor~\cite{van}. The determinant of this unitary matrix still allows the definition of a gauge invariant total Berry phase, so a Chern number can be assigned. Introducing multiband vortex fields in this case is possible~\cite{hat}, but providing a simpler solution without major modifications seems desirable.

In this paper, we want to develop a scheme to determine the joint Chern number of bands that are connected by Dirac cones using single-band vortex fields. Our testing ground will be the Hofstadter model, which can exhibit Dirac cones. Chern numbers of the Hofstadter band structure can be calculated with the Berry formalism of non-degenerate states~\cite{tknn, koh2}, however the bands connected by Dirac cones can only be analyzed implicitly with this approach. The Dirac cones of the Hofstadter model lack a winding number which imposes an additional challenge. The winding in a two-level system is commonly defined by the change of the phase difference of the two components of an eigenstate in reciprocal space around the crossing point (compare e.g. Ref.~\cite{par}). Dirac cones with missing winding number have to be described with winding vectors instead, originally proposed in Ref.~\cite{mon}. We will calculate the Chern numbers of bands in the Hofstadter model connected by Dirac cones through a rotation of the winding vectors in pseudospin space. This gives their Dirac cones the same form as in the common toy Hamiltonian~\cite{mon2}
\begin{equation}\label{the_dirac_hamiltonian}
\hat{H}^{\pm}(\vec{k}) = \hbar v_{\text{F}} (\pm k_{\text{x}} \sigma_\text{x} + k_\text{y} \sigma_\text{y})
\end{equation}
for Dirac fermions with a well-defined winding number of $\pm 1$~\cite{par}. Here, $v_{\text{F}}$ is the Fermi velocity and $\sigma_{\text{x/y}}$ are Pauli matrices in pseudospin space.

This paper is structured as follows. In Sec.~\ref{sec:Vortex_fields} we recapitulate the vortex field formalism for the determination of Chern numbers (Sec.~\ref{subsec:Gapped_bands}), then we consider the issue of topological defects with well-defined winding numbers (Sec.~\ref{subsec:Dirac_matter}) and finally we apply it exemplarily to the Qi–Wu–Zhang (QWZ) model (Sec.~\ref{subsec:Example_qwz}). Sec.~\ref{sec:Hofstadter} is devoted to the Hofstadter model. We show how to treat Dirac cones with non-trivial winding vectors for the Hofstadter model with half a flux quantum per lattice site (Sec.~\ref{subsec:hofstadter_q_2}). Then a scheme for the computation of the Chern numbers for any flux threading of the Hofstadter model, including those with Dirac cones, will be presented (Sec.~\ref{subsec:hofstadter_q_not_2}). In Appendix~\ref{appendix_1} we portray the algebra of magnetic translation operators of the Hofstadter model, the results of which will be used to give some details on calculations concerning the weak coupling limit of the Hofstadter model in Appendices~\ref{appendix_2} and~\ref{appendix_3}. In Appendix~\ref{appendix_4} we explain how the presence of topological defects gives rise to a $\mathbb{Z}_2$ invariant which manifests itself in a discontinuity of the vortex fields.

\section{Vortex fields}\label{sec:Vortex_fields}

\subsection{Patches and vortex fields}\label{subsec:Gapped_bands}

The Chern number of the $n$-th band on the torus $T^2$ comprising the first Brillouin zone is defined as~\cite{sim}
\begin{equation}\label{definition_of_chern_number}
C_{n} = \frac{1}{2\pi} \int_{T^2} \diff \vec{S} \cdot \vec{\nabla}_{\vec{k}} \times \vec{A}^{n}(\vec{k}).
\end{equation}
Here $\diff \vec{S}$ denotes the differential surface vector of the torus and $\vec{A}^{n}(\vec{k})$ is the Berry connection
\begin{equation}\label{Berry_Connection}
\vec{A}^{n}(\vec{k}) = \ci \braket{\Psi^{n}(\vec{k})|\vec{\nabla}_{\vec{k}}|\Psi^{n}(\vec{k})}
\end{equation}
of the $n$-th band with normalized eigenstates $\ket{\Psi^{n}(\vec{k})}$ of a Hamiltonian $\hat{H}(\vec{k})$. Experimentally, $C_{n}$ manifests itself in its relation to the quantum Hall conductance $\sigma_{\text{xy}} = -e^2 C_{n}/h$~\cite{koh, tknn}. We shall have a total of $q$ bands, so $\hat{H}$ is a $q \times q$ matrix. The Berry connection requires eigenstates in reciprocal space to have a well-defined derivative. This is guaranteed by ``fixing the gauge'' locally in $T^2$~\cite{hat}. If we found a suitable continuous ``gauge convention'' over the entire torus, then the Chern number would be equal to zero. This is a consequence of Stokes' theorem, according to which we could rewrite Eq.~(\ref{definition_of_chern_number}) as a line integral over the boundary of the Brillouin zone. On account of periodic boundary conditions in reciprocal space the contributions from opposite edges would necessarily cancel each other. This implies that for every non-trivial Chern insulator ``singular points'' exist: In gapped $q$-band models it is always possible to locally pick smooth normalized eigenstates using a suitable phase convention of the wave function in reciprocal space~\footnote{This is because the projector of eigenstates is smooth, which follows from Eq.~(D1) in~\cite{ans} and the fact that $\hat{H}(\vec{k})$ is only supposed to have analytic functions as matrix elements. This also ensures that the line bundle of the $n$-th band is smooth.}. ``Singularities'', points where the Berry connection diverges, must then originate from the phase convention resulting in a discontinuity of the wave function at some points. With a different phase convention those phases can be made well-defined in the neighborhoods of these points, not on the entire torus.

How do we ``fix the gauge'' and where are the singular points of a given phase convention? Let $\ket{\Phi_{\text{I}}}$ be some normalized state in the Hilbert space. For convenience, we choose $\ket{\Phi_{\text{I}}}$ to be constant, although it is only required to be smooth. Let $S_{\text{I}}$ be the set of points $\vec{k}_{j}$ (which we assume to be discrete~\cite{berry}), where $\braket{\Psi^{n}(\vec{k})|\Phi_{\text{I}}} = 0$. Then we identify smooth eigenstates in $T^2 \setminus S_{\text{I}}$ with a fixed phase convention by projecting $\ket{\Phi_{\text{I}}}$ onto $\ket{\Psi^{n}(\vec{k})}$ with the gauge invariant eigenstate projector $\hat{P}_{\Psi}^{n} = \ket{\Psi^{n}} \bra{\Psi^{n}}$:
\begin{equation}
\ket{\Psi^{n}_{\text{I}}} = \frac{\hat{P}_{\Psi}^{n}(\vec{k})\ket{\Phi_{\text{I}}}}{|\braket{\Psi^{n}(\vec{k})|\Phi_{\text{I}}}|} = \e^{\ci \varphi_{\text{I}}(\vec{k})} \ket{\Psi^{n}(\vec{k})}.
\end{equation}

If a band has a non-trivial Chern number ($C_{n} \neq 0$) then $S_{\text{I}} \neq \emptyset$. At $\vec{k}_{j} \in S_{\text{I}}$ we need a second phase convention, where corresponding eigenstates $\ket{\Psi^{n}_{\text{II}}(\vec{k})}$ are smooth at $\vec{k}_{j}$. A neighborhood of $\vec{k}_{j}$ is denoted as a patch $P_j$. Within the patches $P = \bigcup_{j} P_{j}$ a different phase convention is set up with some other normalized state $\ket{\Phi_{\text{II}}}$
\begin{equation}
\ket{\Psi^{n}_{\text{II}}(\vec{k})} = \frac{\hat{P}_{\Psi}^{n}(\vec{k})\ket{\Phi_{\text{II}}}}{|\braket{\Psi^{n}(\vec{k})|\Phi_{\text{II}}}|} = \e^{\ci \varphi_{\text{II}}} \ket{\Psi^{n}(\vec{k})},
\end{equation}
where $\lbrace P_j \rbrace$ and $\ket{\Phi_{\text{II}}}$ are chosen such that the condition $\braket{\Psi^{n}(\vec{k})|\Phi_{\text{II}}} \neq 0$ is fulfilled for all $\vec{k} \in P$. In turn there must be a set of points $S_{\text{II}} \subset T^2 \setminus P$, where $\braket{\Psi^{n}(\vec{k})|\Phi_{\text{II}}} = 0$.

For $\vec{k} \in P \setminus S_{\text{I}}$ the wave functions $\ket{\Psi_{\text{I}}^{n}(\vec{k})}$ and $\ket{\Psi_{\text{II}}^{n}(\vec{k})}$ can be related by the transition function $\e^{\ci \chi(\vec{k})}$:
\begin{equation}
\ket{\Psi^{n}_{\text{I}}(\vec{k})} = \ket{\Psi^{n}_{\text{II}}(\vec{k})} \e^{\ci ( \varphi_{\text{I}} - \varphi_{\text{II}})} = \ket{\Psi^{n}_{\text{II}}(\vec{k})} \e^{\ci \chi (\vec{k})}
\end{equation}
and as a result their Berry connections are related like
\begin{equation}\label{relation_of_connections}
\vec{A}^{n}_{\text{I}} = \vec{A}^{n}_{\text{II}} + \ci \e^{-\ci \chi} \vec{\nabla}_{\vec{k}} \e^{\ci \chi}.
\end{equation}
Note that we can vary the size of patches $P$, so we can define $\chi(\vec{k})$ everywhere except at $\vec{k} \in S_{\mathrm{I}}\cup S_{\mathrm{II}}$.

Now Stokes' theorem can be applied and results in line integrals along the oriented boundary of the patches $\partial P$ (provided $\partial P \cap S_{\text{II}} = \emptyset$)~\footnote{In mathematical terms the function $\e^{\ci \chi}$ is a transition function, which connects the two locally defined sections $\ket{\Psi^{n}_{\text{I}}}$ and $\ket{\Psi^{n}_{\text{II}}}$ of the line bundle of the $n$-th band. The transition function contains the information how ``twisted'' the line bundle is, i.e. the Chern number. Usually this transformation is called ``gauge transformation'' in the literature, even though $\ci\e^{-\ci \chi} \diff \e^{\ci \chi}$ is not an exact differential form. It does however leave the equations of motion invariant, as, analogously to Dirac's magnetic monopole argument~\cite{god}, the Aharonov–Bohm phase of this transformation must not be measurable (which in turn explains, why $C_{n}$ has to be an integer).}
\begin{equation}\label{Chern_number_winding}
C_n = \frac{1}{2\pi} \oint_{\partial P} \diff \vec{k} \left( \vec{A}^{n}_{\text{II}} - \vec{A}^{n}_{\text{I}} \right) = \frac{1}{2\pi} \oint_{\partial P} \diff \Arg(\e^{\ci\chi}).
\end{equation}
We have to extract from Eq.~(\ref{Chern_number_winding}) how many times the phase $\chi$ winds around each patch in which direction. This can be facilitated with the introduction of vortex fields: We plot a vector field in polar coordinates with
\begin{equation}
\Arg(\e^{\ci\chi}) = \Arg(\braket{\Phi_{\text{II}}|\Psi^{n}_{\text{I}}})
\end{equation}
as the azimuth and $|\braket{\Phi_{\text{II}}|\Psi^{n}_{\text{I}}}|$ as the radius (see e.g. Fig.~\ref{fig:QWZ}; to make vortices easier to spot, the azimuth is further represented by the color of the vector field). Then calculating $C_n$ is equivalent to counting all the vortices, where the radius does not vanish and where the vortex field is not continuously differentiable. These vortices are associated with flux tubes of monopole charges (see Sec.~\ref{subsec:Dirac_matter}), so we call them ``flux carrying vortices''~\footnote{We want to add that there is an interesting relation between the origin of these discontinuities in the vortex field and the fact~\cite{che} that one can also determine the Chern number by calculating the algebraic sum of zeros of a smooth, holomorphic section of the line bundle in question: It is the restraint of having normalized wave functions that leads to discontinuities---flux carrying vortices---at these former zero points.}. There are also vortices where $\braket{\Phi_{\text{II}}|\Psi^{n}_{\text{I}}}$ does vanish. These ``trivial vortices'' must not be summed up to calculate the Chern number as the field is smooth at the vortex centers.

What is a convenient choice for the phase convention? If $\ket{\Phi_{\text{I}}}$ and $\ket{\Phi_{\text{II}}}$ are the first and second Cartesian unit vectors of the $q$-dimensional Hilbert space of the parameterized Hamiltonian matrix, then the vortex field is equivalent to the second component of the eigenstates $\ket{\Psi^{n}_{\text{I}}}$ whose first component is non-negative real if $S_{\text{I}} \cap S_{\text{II}} = \emptyset$. In the following we will use this phase convention unless stated otherwise.

\subsection{Topological defects}\label{subsec:Dirac_matter}

\begin{figure*}
\subfloat[]{
  \includegraphics[height=0.35\textwidth]{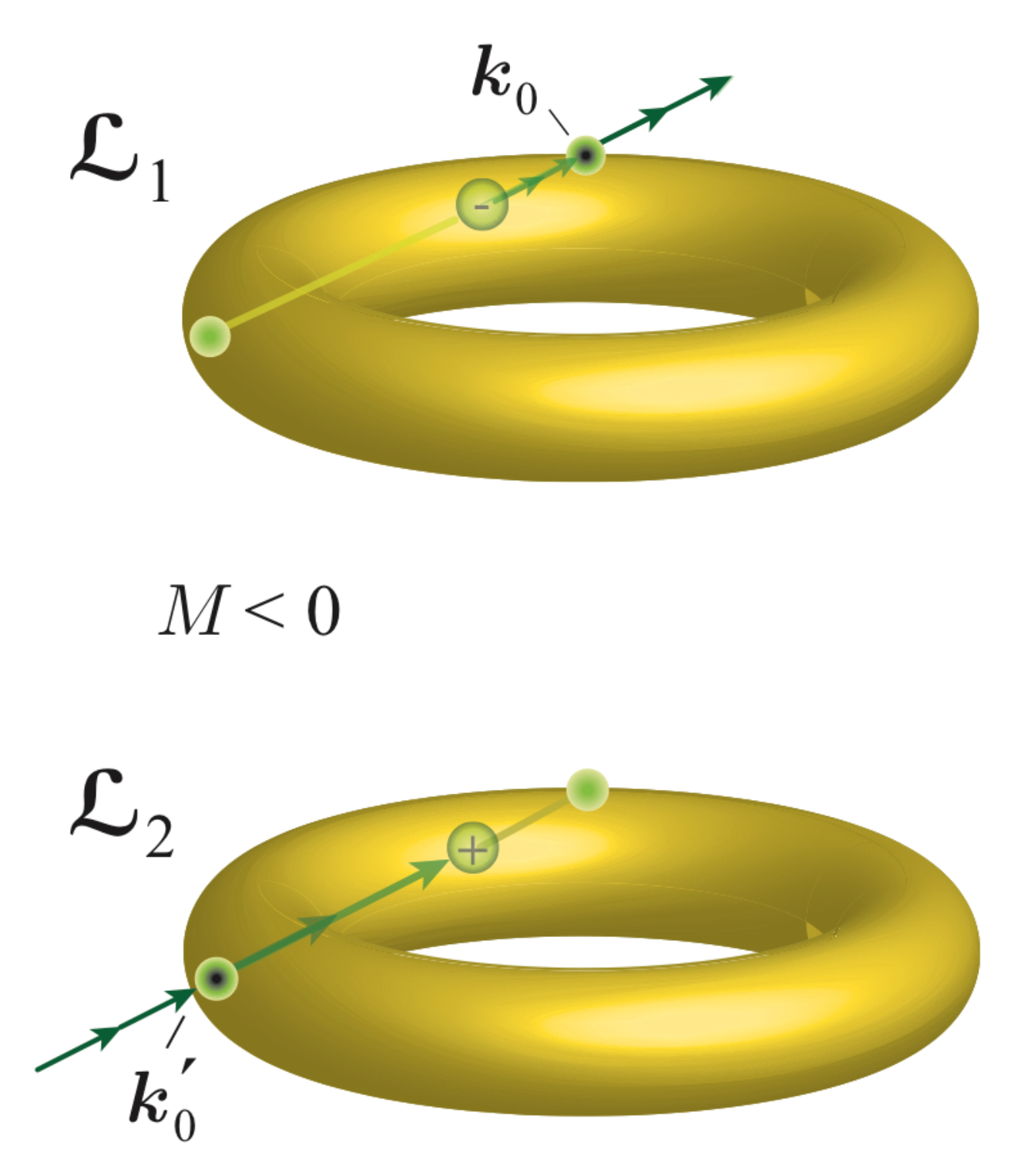}
  \label{subfig:torus_smaller}
}\hfil
\subfloat[]{
  \includegraphics[height=0.35\textwidth]{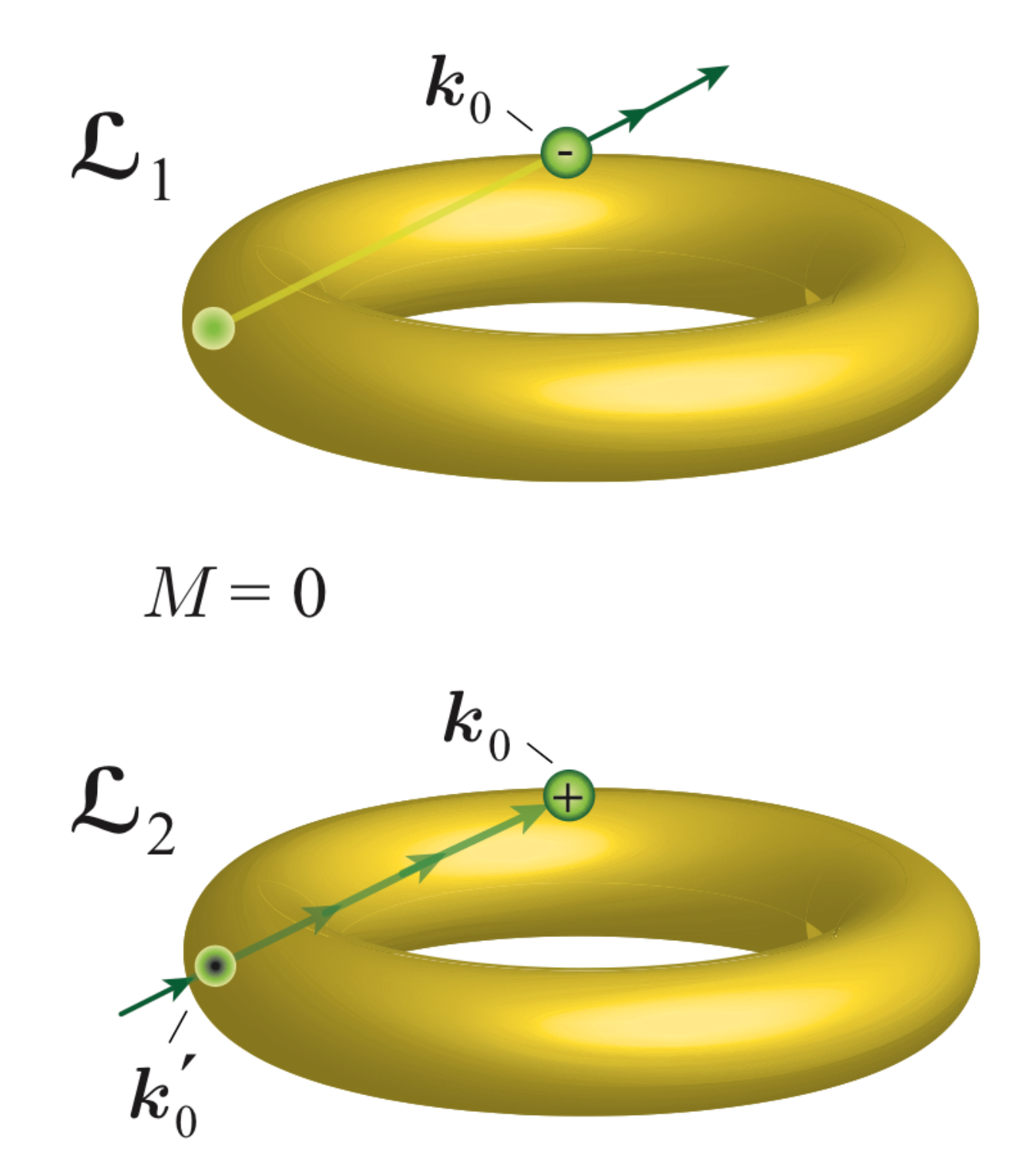}
  \label{subfig:torus_equal}
}\hfil
\subfloat[]{
  \includegraphics[height=0.35\textwidth]{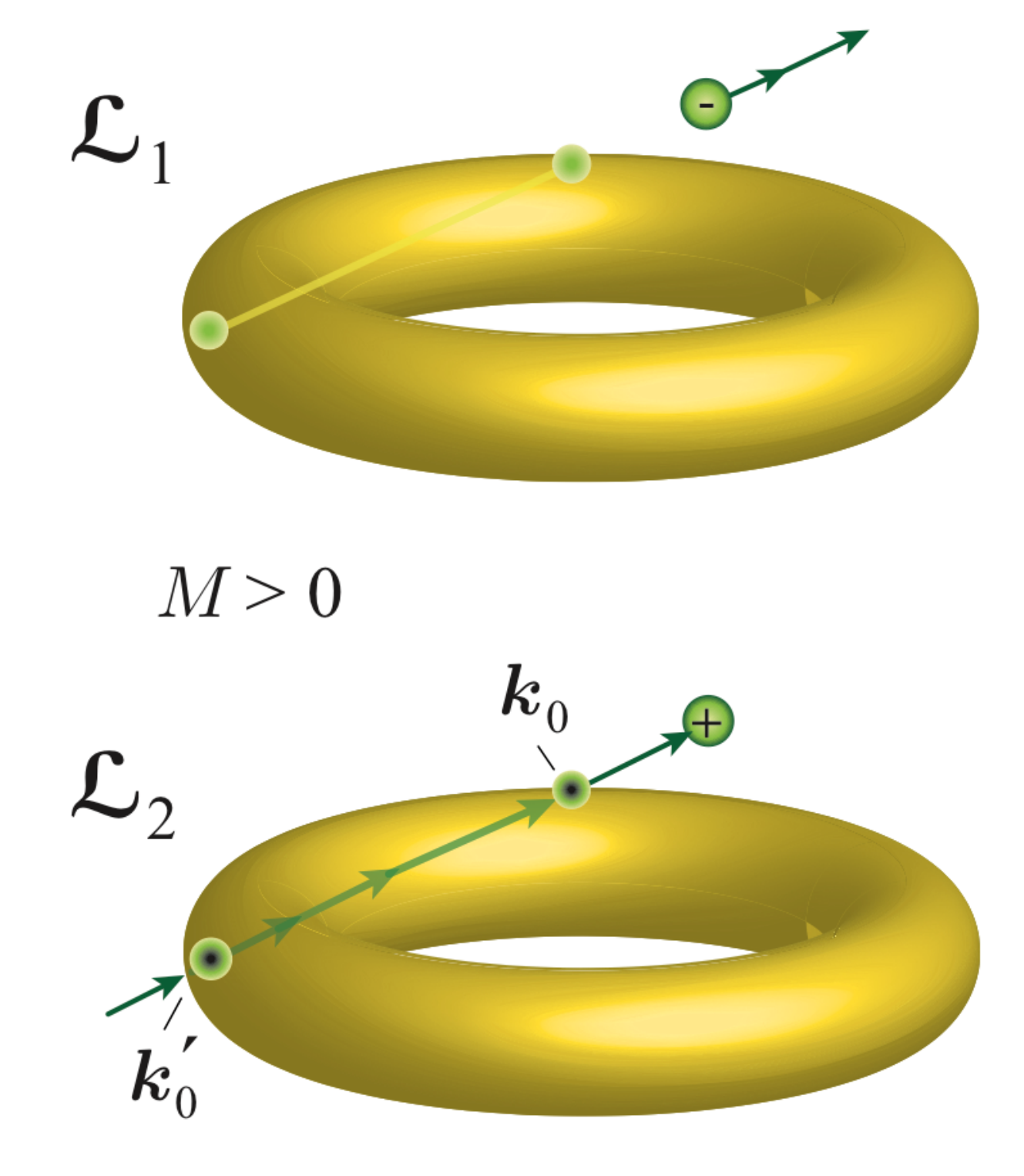}
  \label{subfig:torus_larger}
}
\caption{Monoples in a torus geometry for the eigenspaces $\mathcal{L}_1$ and $\mathcal{L}_2$ for (a) $M < 0$, (b) $M = 0$ and (c) $M > 0$. The flux tubes are represented by the solid line with arrows. They pierce the surface of the torus at $\vec{k}_0$ and $\vec{k}_0'$ and at these points there will be flux carrying (anti)vortices as a consequence; otherwise there will be trivial vortices identified by yellow lines piercing the surface.}
\label{fig:tori}
\end{figure*}

It is instructive to include certain two-dimensional topological defects---Dirac cones---into the analysis. They correspond to Weyl points, topological defects that exist in three dimensions. They are responsible for the emergence of flux carrying vortices. It is well known that Chern numbers can be related to monopole charges in a solid torus parametrization of the Hamiltonian matrix~\cite{sim}. Let $\lbrace \mathcal{L}_{1}, \dots, \mathcal{L}_{q} \rbrace$ be the eigenspaces of a $q \times q$ Hamiltonian of which all eigenvalues are gapped for any $\vec{k}$. In mathematical terms, these eigenspaces correspond to line bundles, so that their direct sum makes up the entire Bloch bundle of the Hamiltonian~\cite{kau}. Let $k_{\text{x}}, k_{\text{y}}, r$ be a parametrization of the solid torus, where $r$ is a new radial parameter of the Hamiltonian. The original problem is then given for a fixed $r = R$. Then eigenstates $\Psi_{r}^{j}(\vec{k})$ belonging to the eigenspace $\mathcal{L}_{j}$ can locally be defined uniquely, except at certain points $(\vec{k}_{j}, r_{j})$, $r_{j} \neq R$ inside the torus, where the $j$-th band is degenerate either with the $(j+1)$-th or $(j-1)$-th band. These are the points, where the monopole charges are located. The presence of monopole charges implies the existence of flux tubes for the same reason as in the magnetic monopole geometry~\cite{god}. Flux tubes manifest themselves as the flux carrying vortices in the vortex field frame. This can be shown by calculating their Aharonov–Bohm phase. Assuming that at some point of the Brillouin zone there is a flux carrying vortex, we can identify, using Eq.~(\ref{relation_of_connections})
\begin{equation}\label{flux_tube_is_vortex}
\begin{aligned}[b]
\oint_{C_{\epsilon}} \vec{A}_{\text{I}}^{n} \diff \vec{k}
&= \oint_{C_{\epsilon}} (\vec{A}_{\text{II}}^{n} + \ci \e^{-\ci \chi} \vec{\nabla}_{\vec{k}} \e^{\ci \chi}) \diff \vec{k} \\
&= - \oint_{C_{\epsilon}} \diff \Arg(\e^{\ci \chi}),
\end{aligned}
\end{equation}
where $C_{\epsilon}$ is an infinitesimal contour around the vortex. The last step of Eq.~(\ref{flux_tube_is_vortex}) follows, because inside $C_{\epsilon}$ $\vec{A}_{\text{II}}$ has no singularity. We see that the flux is the negative of the winding number, because the flux of the flux tube flows from outside the torus to the monopole, from where flux then flows through the surface of the torus, or vice versa for negative monopole charges.

A monopole located at the surface of the torus manifests itself as a Dirac point that connects two bands via an eigenenergy degeneracy. We cannot define a Chern number for any of the two connected bands individually, only for both bands jointly. The eigenenergies and eigenstates of $\hat{H}^{+}$ of Eq.~(\ref{the_dirac_hamiltonian}) are given by~\cite{tka}
\begin{equation}\label{eigenstates_dirac_cone}
\begin{aligned}
E_s &= s \sqrt{k_{\text{x}}^2 + k_{\text{y}}^2 + M^2} \\
\ket{\Psi_{\text{I}}^{s}(\vec{k})} &= \frac{1}{\sqrt{2}} \begin{pmatrix}
\sqrt{1 + s u_{\text{z}}} \\
s \e^{\ci\phi} \sqrt{1 - su_{\text{z}}}
\end{pmatrix},
\end{aligned}
\end{equation}
with $u_{\text{z}} = M / E_+$ and $\tan \phi = k_{\text{y}} / k_{\text{x}}$. The index $s = \pm$ denotes the band. $\hat{H}^{+}$ describes a Dirac cone with positive winding number. This is evident because the phase difference $\phi$ between the first and second component changes by $+2\pi$ when moving counterclockwise around the contact point at $\vec{k} = 0$. Remark that in the chosen phase conventions for the vortex field this phase difference equals the phase of the vortex field. Also note that $\lbrace \vec{k} \rbrace$ is supposed to cover the Brillouin zone, which is a compact manifold; it appears to be unbound in Eq.~(\ref{eigenstates_dirac_cone}), as there we only consider the vicinity of a Dirac point.

We can break the Dirac cone up and move the monopole charge inside or outside the torus with a ``mass'' $M \sigma_{\text{z}}$ as illustrated in Fig.~\ref{fig:tori}. In the gauge choice of Eq.~(\ref{eigenstates_dirac_cone}) we get a flux carrying vortex in the lower band when $M > 0$ and we get a flux carrying vortex in the upper band for $M < 0$. Note that for the states of Eq.~(\ref{eigenstates_dirac_cone}) we address only the vicinity of the point $\vec{k}_0$ in Fig.~\ref{fig:tori}. We assume $|M|$ to be sufficiently small, so that the change of the eigenstates at $\vec{k}$ away from the Dirac point is negligible. The change of the number of flux carrying vortices in each of the two vortex fields corresponds to a change of the number of associated flux tubes and therefore the number of monopoles in the tori of the two eigenspaces $\mathcal{L}_{1}$ and $\mathcal{L}_2$. Upon breaking up the Dirac cone of Eq.~(\ref{eigenstates_dirac_cone}), characterized by a positive winding number, a flux carrying vortex is generated in both cases, $M<0$ or $M>0$, in one of the bands. Hence, in this phase convention its contribution to the total Chern number of the two bands is always $+1$. We can therefore still work with single-band vortex fields: We plot the vortex fields of both bands in this gauge, count all flux carrying vortices away from the Dirac points, then add the contributions from the Dirac cones and find the joint Chern number of both bands. Note that this is not a contradiction to the result of Ref.~\cite{sim}, which states that a monopole resulting from the degeneracy of two bands will not yield any contribution to the overall Chern number of both bands. In contrast, we determine the Chern number by counting flux tubes in a distinct phase convention, which is a different concept than that used in Ref.~\cite{sim}.

\subsection{Qi–Wu–Zhang model}\label{subsec:Example_qwz}

\begin{figure}[tb]
\subfloat[]{
  \includegraphics[width=0.33\textwidth]{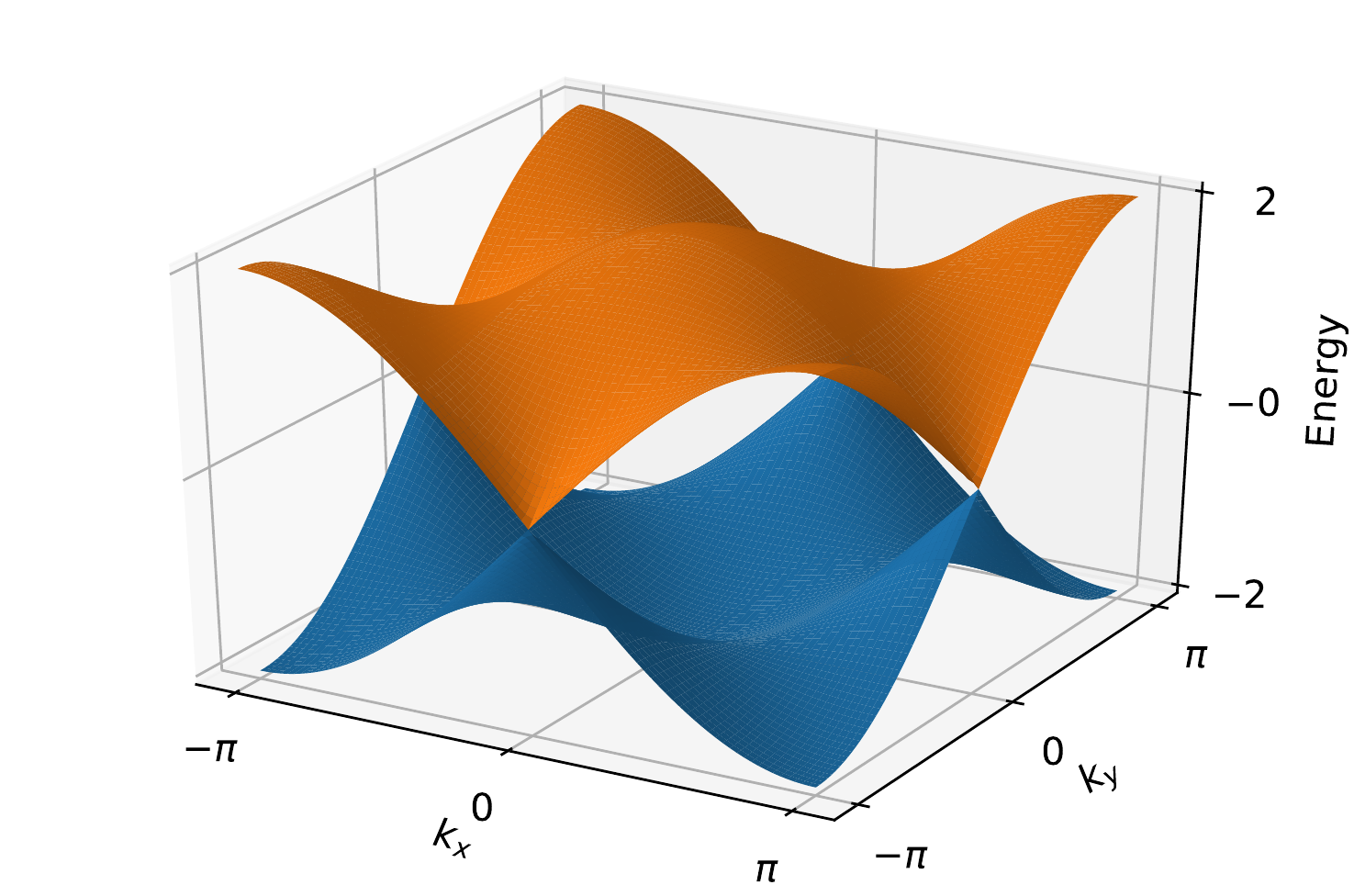}
}\hfil
\subfloat[]{
  \includegraphics[width=0.33\textwidth]{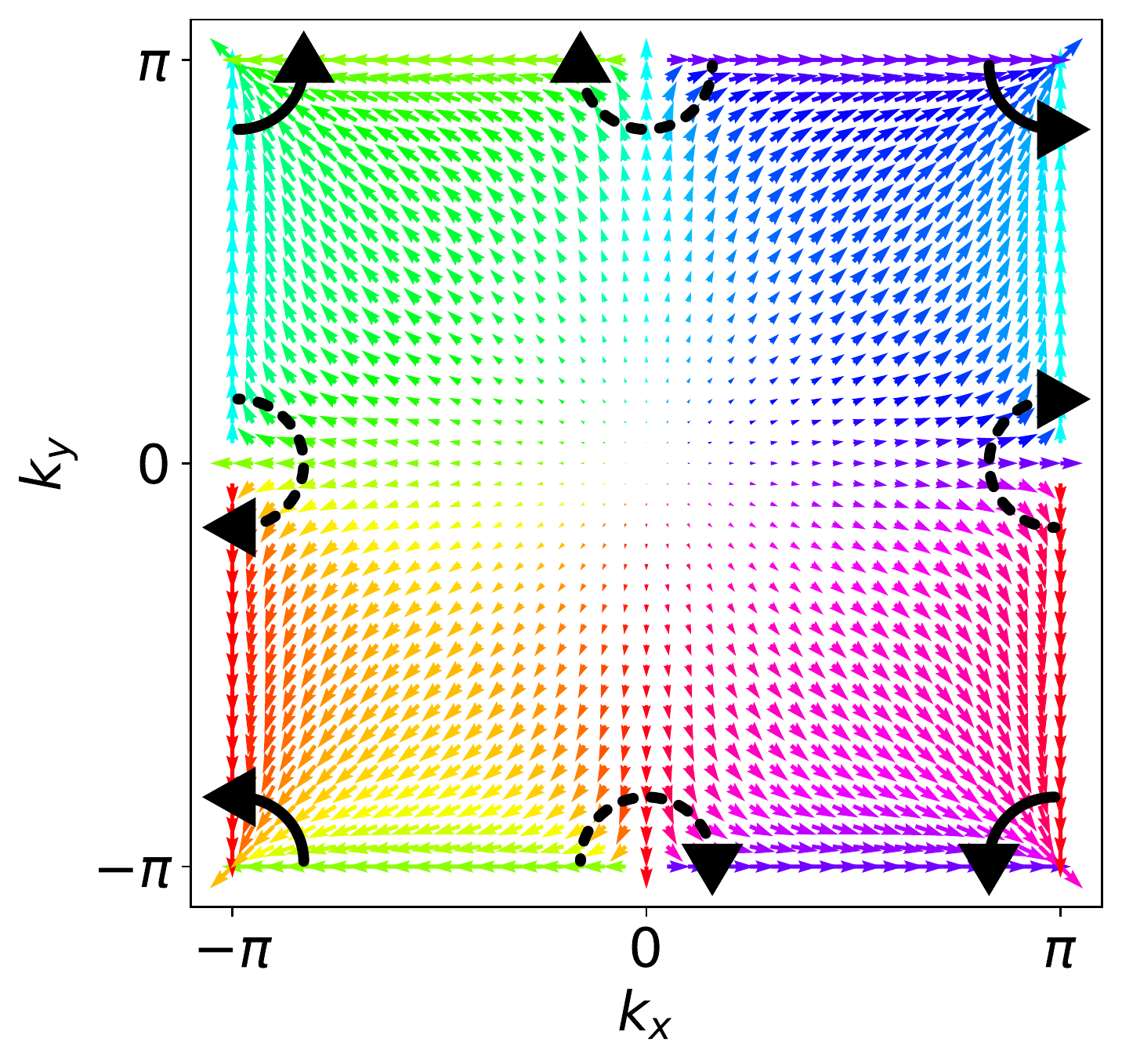}
}\hfil
\subfloat[]{
  \includegraphics[width=0.33\textwidth]{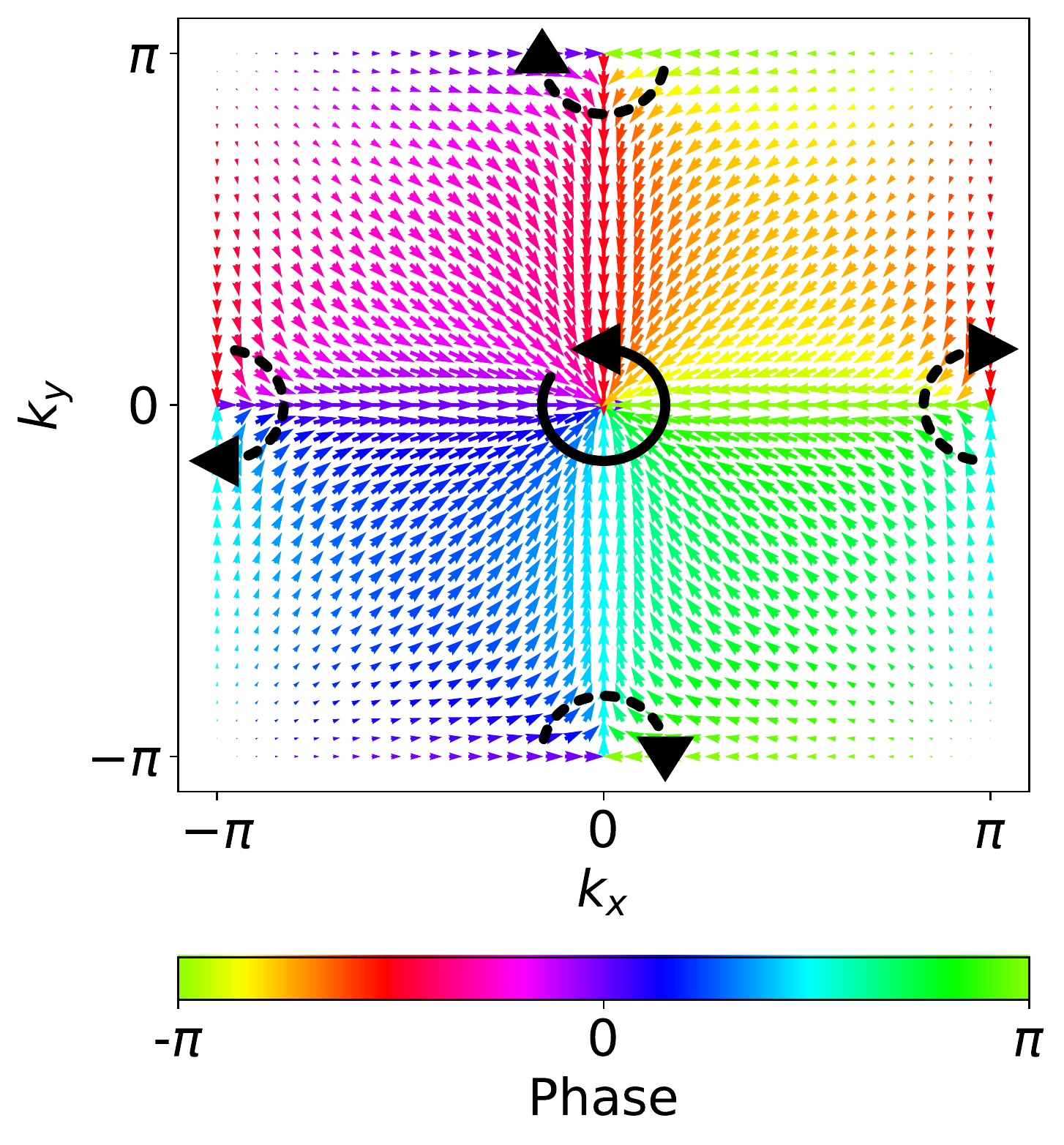}
}
\caption{QWZ model for $u = 0$. (a) Energy spectrum, (b) Vortex field of the upper band and (c) Vortex field of the lower band. Vortices are marked as oriented arrows, Dirac cones as dotted oriented arrows.}
\label{fig:QWZ}
\end{figure}

As an example for the determination of the Chern number $C$ from the vortex field we study the QWZ model~\cite{qwz, asb}. The Hamiltonian reads
\begin{equation}
\hat{H}_{\text{QWZ}} = \sin k_{\text{x}} \sigma_{\text{x}} + \sin k_{\text{y}} \sigma_\text{y} + (u + \cos k_\text{x} + \cos k_\text{y})\sigma_\text{z}.
\end{equation}
The general solution of the eigenvectors of a Hamiltonian $\hat{H} = \vec{f} \vec{\sigma}$ with Pauli matrices $\vec{\sigma} = (\sigma_\text{x}, \sigma_\text{y}, \sigma_\text{z})^{T}$ and eigenenergies $E_{\pm} = \pm \sqrt{\sum_{j} f_j^2}$ is
\begin{equation}\label{general_2_times_2}
\ket{\Psi_{\text{I}}^{\pm}} = \frac{1}{\sqrt{2}} \frac{1}{\sqrt{E_{+}^2 \pm f_\text{z} E_{+}}} \begin{pmatrix}
E_{+} \pm f_\text{z}\\
\pm(f_\text{x} + \ci f_\text{y})
\end{pmatrix},
\end{equation}
so vortices can only emerge at $f_\text{x} = f_\text{y} = 0$. Therefore, for the QWZ model only the time reversal invariant momentum (TRIM) points are relevant for our purposes. For example in linear approximation around $(\pi,0)$ and $(0, \pi)$ the effective Hamiltonians are
\begin{equation}
\begin{gathered}
\hat{H}^{(\pi, 0)}_{\text{QWZ}} \approx - (k_\text{x} - \pi) \sigma_\text{x} + k_\text{y} \sigma_\text{y} + u \sigma_\text{z} \\
\hat{H}^{(0, \pi)}_{\text{QWZ}} \approx k_\text{x} \sigma_\text{x} - (k_\text{y} - \pi) \sigma_\text{y} + u \sigma_\text{z},
\end{gathered}
\end{equation}
which both yield a flux carrying antivortex in the upper band for $u<0$, otherwise a flux carrying vortex in the lower band for $u>0$ as presented in Sec.~\ref{subsec:Gapped_bands}. For $u=0$ we find Dirac cones each with a contribution of $-1$ to the total Chern number of both bands, see Fig.~\ref{fig:QWZ}. Similar analyses can be carried out for other TRIM points. Adding the contributions together we can determine the Chern number of the upper band depending on $u$ (similar to Ref.~\cite{asb}):
\begin{equation}
\begin{aligned}
C = 0 \quad &: \quad u<-2\\
C = -1 \quad &: \quad -2 < u <0\\
C = 1 \quad &: \quad 0 < u < 2\\
C = 0 \quad &: \quad 2 < u
\end{aligned}
\end{equation}
At $u=-2,0,2$ the band gap vanishes due to Dirac cones. If the band gap closes, it appears as if there were a flux carrying vortex in both vortex fields at the same point. Fig.~\ref{fig:QWZ} shows energy spectrum and vortex fields for $u=0$. Dirac cones emerge at $(0, \pi)$ and $(\pi, 0)$. They have counterclockwise winding and therefore contribute $-2$ to the Chern number. Together with the vortex at $(\pi, \pi)$ in the upper band and the vortex at $(0, 0)$ in the lower band we get a total Chern number $C=0$ as it should be for the total Chern number of a multiband system. This is a result of the fact that we can write the sum of the line bundles of a Hermitian $q \times q$ Hamiltonian as $T^2 \times \mathbb{C}^{q}$, which is trivial~\cite{sim, kau}.

\section{Hofstadter model}\label{sec:Hofstadter}

The Hofstadter Hamiltonian exhibits Dirac cones that break the usual division of having clockwise or counterclockwise winding. The Hofstadter Hamiltonian is a tight-binding model on a square lattice in an external magnetic field~\cite{hof}:
\begin{equation}\label{real_space_hofstadter}
\begin{aligned}[b]
\hat{H} =& -t_a \displaystyle\sum_{m, n} \displaystyle\sum_{\mu = 1}^{q} \left( c_{m,n}^{\mu + 1} \right)^{\dagger} c_{m, n}^{\mu} + \text{h.c.} \\
&-t_b \displaystyle\sum_{m, n} \displaystyle\sum_{\mu = 1}^{q} \left( c_{m,n+1}^{\mu} \right)^{\dagger} c_{m, n}^{\mu} \e^{-2\pi \ci \varphi \mu} + \text{h.c.},
\end{aligned}
\end{equation}
where $(m,n)$ is the position of the (magnetic) unit cell, $\mu$ a sublattice index, $t_a$ and $t_b$ are hopping parameters, and $\varphi = p/q$. The additional phase factors (in comparison to the field free case) break the translation symmetry of the square lattice, requiring the unit cells to consist of $q$ lattice sites. They are the Peierls phases~\cite{lut, hof} and can be conceptualized as Aharonov–Bohm phases originating from a flux piercing through each plaquette of the square lattice. They can be calculated by replacing the field-free hopping terms
\begin{equation}\label{peierls_phase}
c^{\dagger}_{\vec{R} + \vec{e}_{\text{x/y}}} c^{\pdagger}_{\vec{R}} \to c^{\dagger}_{\vec{R} + \vec{e}_{\text{x/y}}} c^{\pdagger}_{\vec{R}} \e^{-\ci \frac{e}{\hbar c} \int_{\vec{R}}^{\vec{R} + \vec{e}_{\text{x/y}}} \vec{A}(\vec{r}') \diff \vec{r}'}
\end{equation}
and depend on the choice of $\vec{A}$. We picked $\vec{A} = Bx\vec{e}_\text{y}$. The flux per lattice site is $\varphi = p/q = B/\Phi_0$ in units of the flux quantum $\Phi_0$, and $B$ is the external magnetic field (the lattice constants are set equal to one). The explicit inclusion of a sublattice is chosen to avoid the unusual Fourier transformation sometimes applied by authors, e.g. in~\cite{koh2, ber, fra}. There, a Brillouin zone of dimensions $-\pi \leq k_\text{x}, k_\text{y} \leq \pi$ is defined and then split up into $q$ sections that serve as degrees of freedom instead of working with proper sublattice indices. We block diagonalize Eq.~(\ref{real_space_hofstadter}) with
\begin{equation}
c_{k_\text{x}, k_\text{y}}^{\mu} = \sqrt{\frac{1}{N}} \displaystyle\sum_{m,n}\e^{-\ci k_\text{x} q m} \e^{-\ci k_\text{y} n} c_{m,n}^{\mu},
\end{equation}
where $N$ is the number of unit cells. For convenience, we introduce the abbreviation $c_{\mu} = c_{k_\text{x}, k_\text{y}}^{\mu}$. Note the continuation condition $c_{\mu + q} = \e^{\ci k_\text{x} q} c_{\mu}$ obtained from $c_{m,n}^{\mu + q} = c_{m+1, n}^{\mu}$. The Hamiltonian then becomes
\begin{equation}
\hat{H} = \displaystyle\sum_{k_\text{x} = - \pi/q}^{\pi/q}\displaystyle\sum_{k_\text{y} = -\pi}^{\pi} \hat{h}_{k_\text{x}, k_\text{y}},
\end{equation}
with
\begin{equation}\label{h_kx_ky}
\begin{aligned}[b]
\hat{h}_{k_\text{x}, k_\text{y}} = \displaystyle\sum_{\mu = 1}^{q} -&t_a c_{\mu + 1}^{\dagger} c^{\pdagger}_{\mu} - t_a c_{\mu}^{\dagger} c^{\pdagger}_{\mu + 1} \\
-& 2t_b \cos(k_\text{y} + 2\pi \varphi \mu) c_{\mu}^{\dagger} c^{\pdagger}_{\mu}
\end{aligned}
\end{equation}
or written as a matrix $h(k_\text{x}, k_\text{y})$ with the elements $h_{\mu, \mu'} = \braket{\mu|\hat{h}_{k_\text{x}, k_\text{y}}|\mu'}$, $\ket{\mu} = c_{\mu}^{\dagger} \ket{0}$
\begin{equation}\label{hofstadter_matrix}
h = \begin{pmatrix}
v_1 & -t_a & \scalebox{1.5}{0} & -t_a \e^{-\ci qk_\text{x}} \\
-t_a & v_2 & \ddots & \scalebox{1.5}{0} \\
\scalebox{1.5}{0} & \ddots & \ddots & -t_a \\
-t_a \e^{\ci qk_\text{x}} & \scalebox{1.5}{0} & -t_a & v_q
\end{pmatrix}.
\end{equation}
Here we have defined $v_{\mu} = -2t_b \cos(k_\text{y} + 2\pi \varphi \mu)$. Note that the matrix in Eq.~(\ref{hofstadter_matrix}) is $2\pi$-periodic in $k_\text{y}$ and $2\pi/q$-periodic in $k_\text{x}$ as it should be considering the shape of our unit cell originating from the choice of $\vec{A}$. The notation from Ref.~\cite{koh2} leads to $k_\text{x}$ and $k_\text{y}$ being swapped in the matrix, which contradicts the definition of the Brillouin zone. Then defining patches over the boundary conditions of the actual Brillouin zone cannot be done.

\subsection{\boldmath{$p/q = 1/2$}: Non-trivial winding vectors}\label{subsec:hofstadter_q_2}

\begin{figure}
  \centering
  \includegraphics[width=1.0\linewidth]{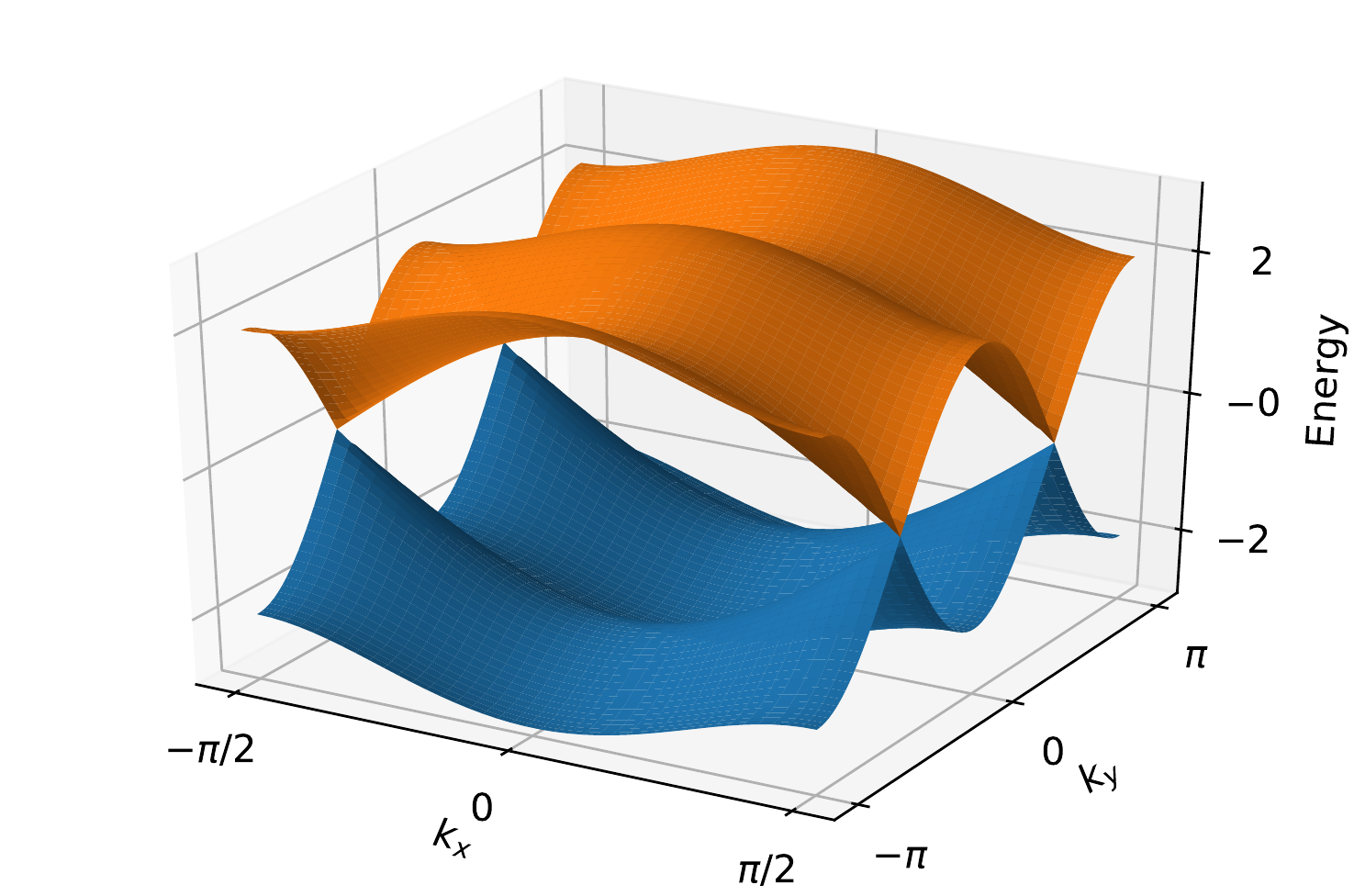}
  \caption{Energy spectrum of the Hofstadter model for $p/q=1/2$.}\label{fig:E_2}
\end{figure}

\begin{figure*}
\subfloat[]{
  \includegraphics[height=0.35\textwidth]{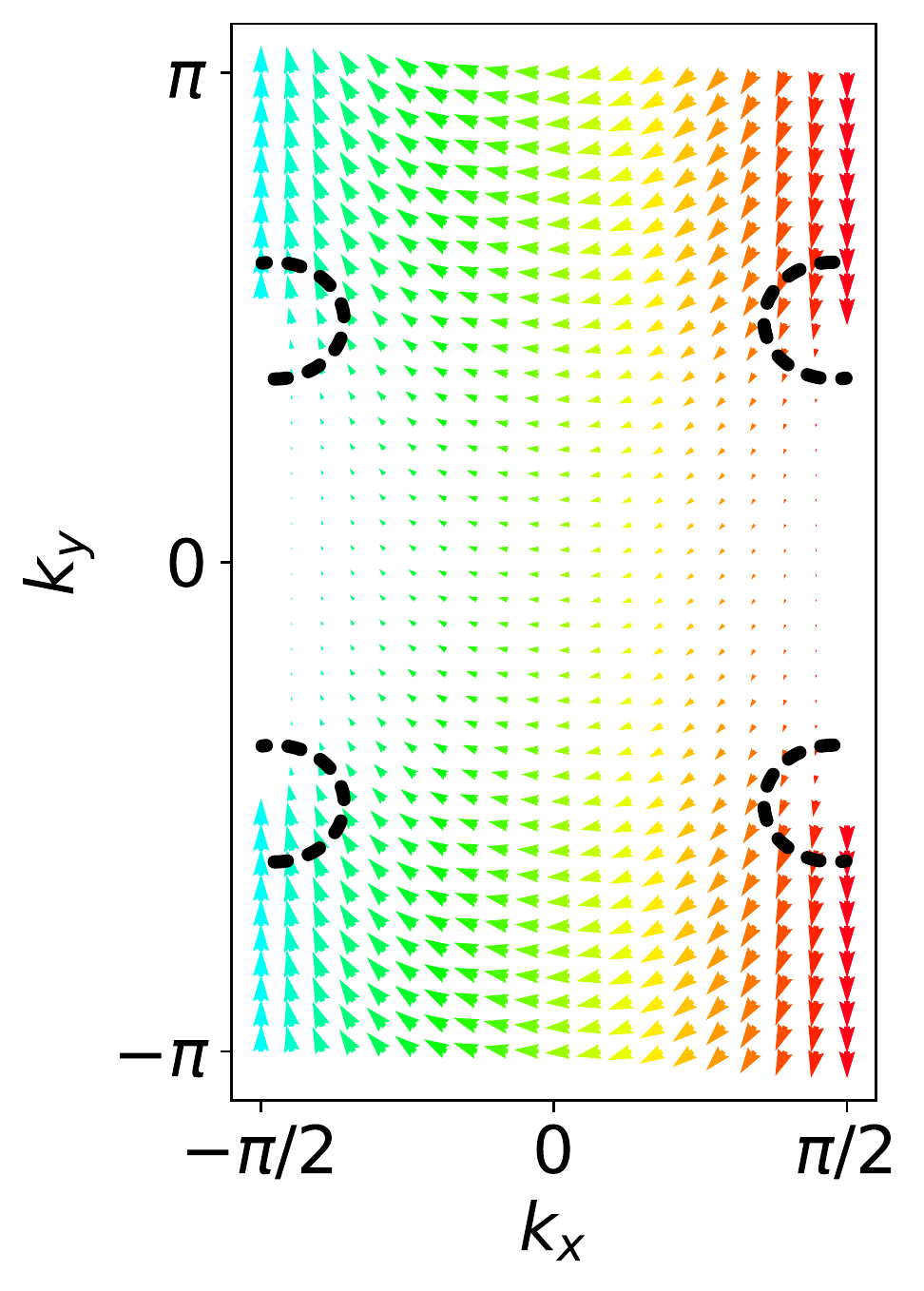}
  \label{subfig:q_2_top}
}\hfil
\subfloat[]{
  \includegraphics[height=0.35\textwidth]{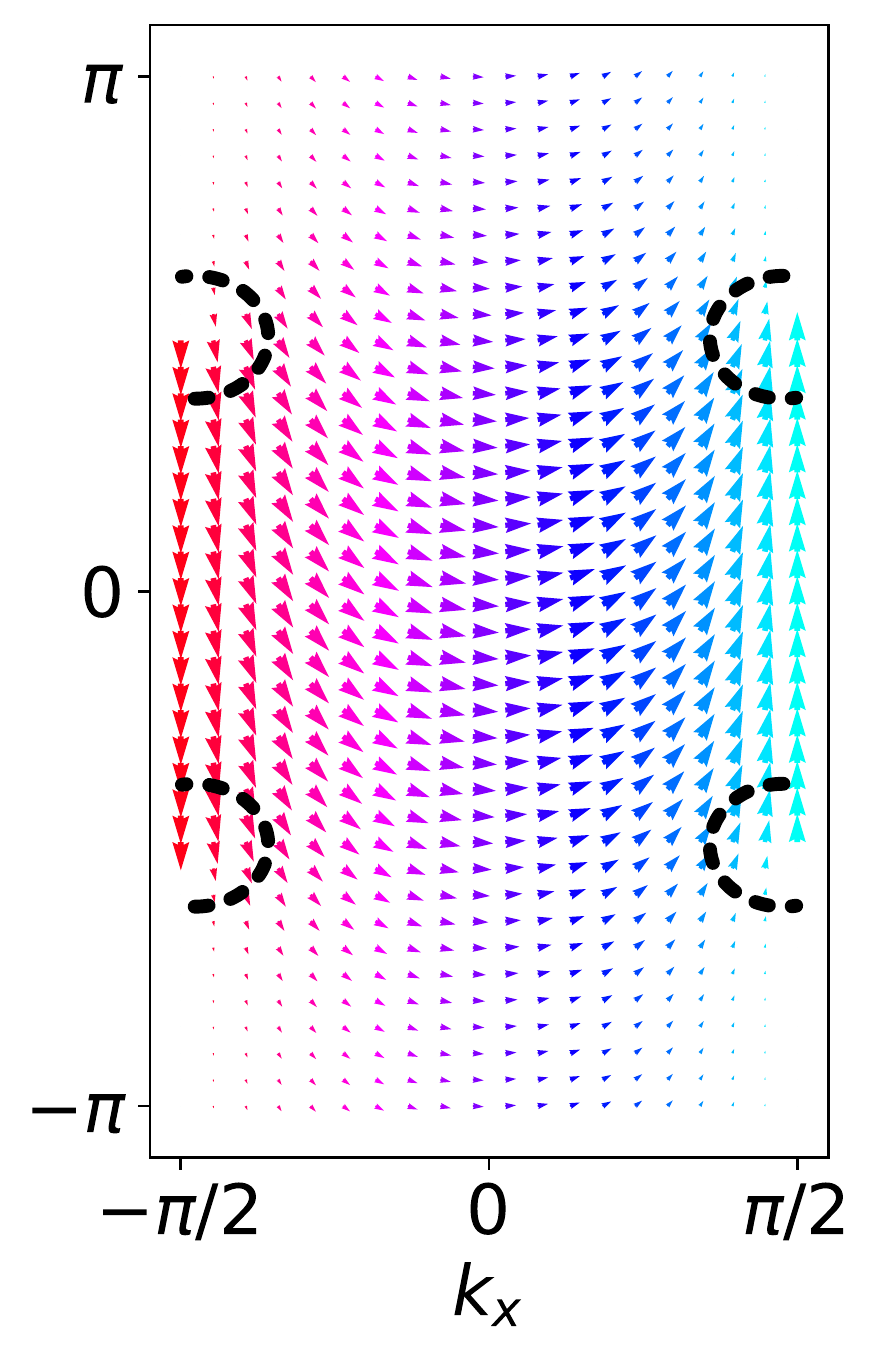}
  \label{subfig:q_2_sub}
}\hfil
\subfloat[]{
  \includegraphics[height=0.35\textwidth]{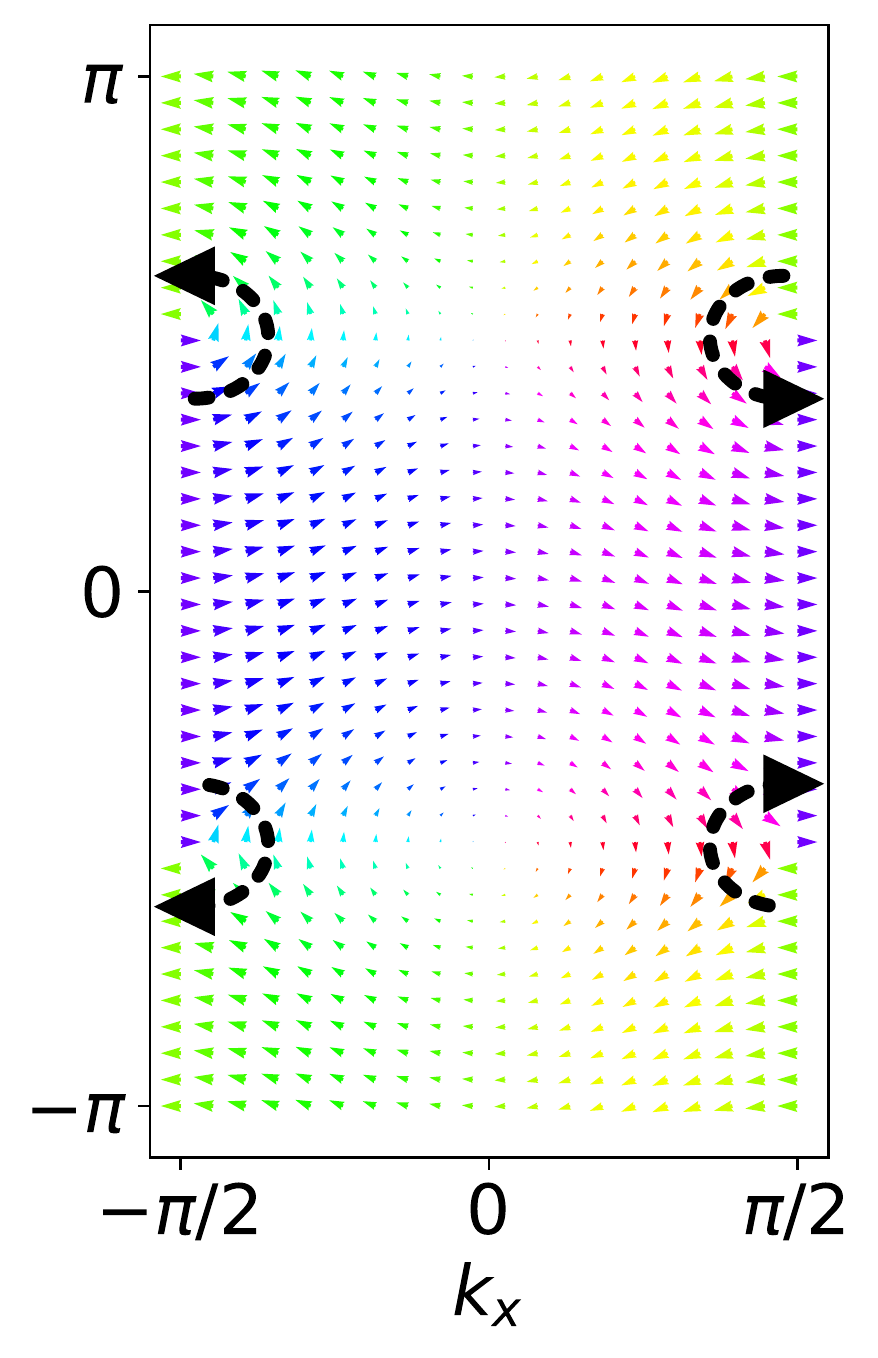}
  \label{subfig:q_2_rot_top}
}\hfil
\subfloat[]{
  \includegraphics[height=0.35\textwidth]{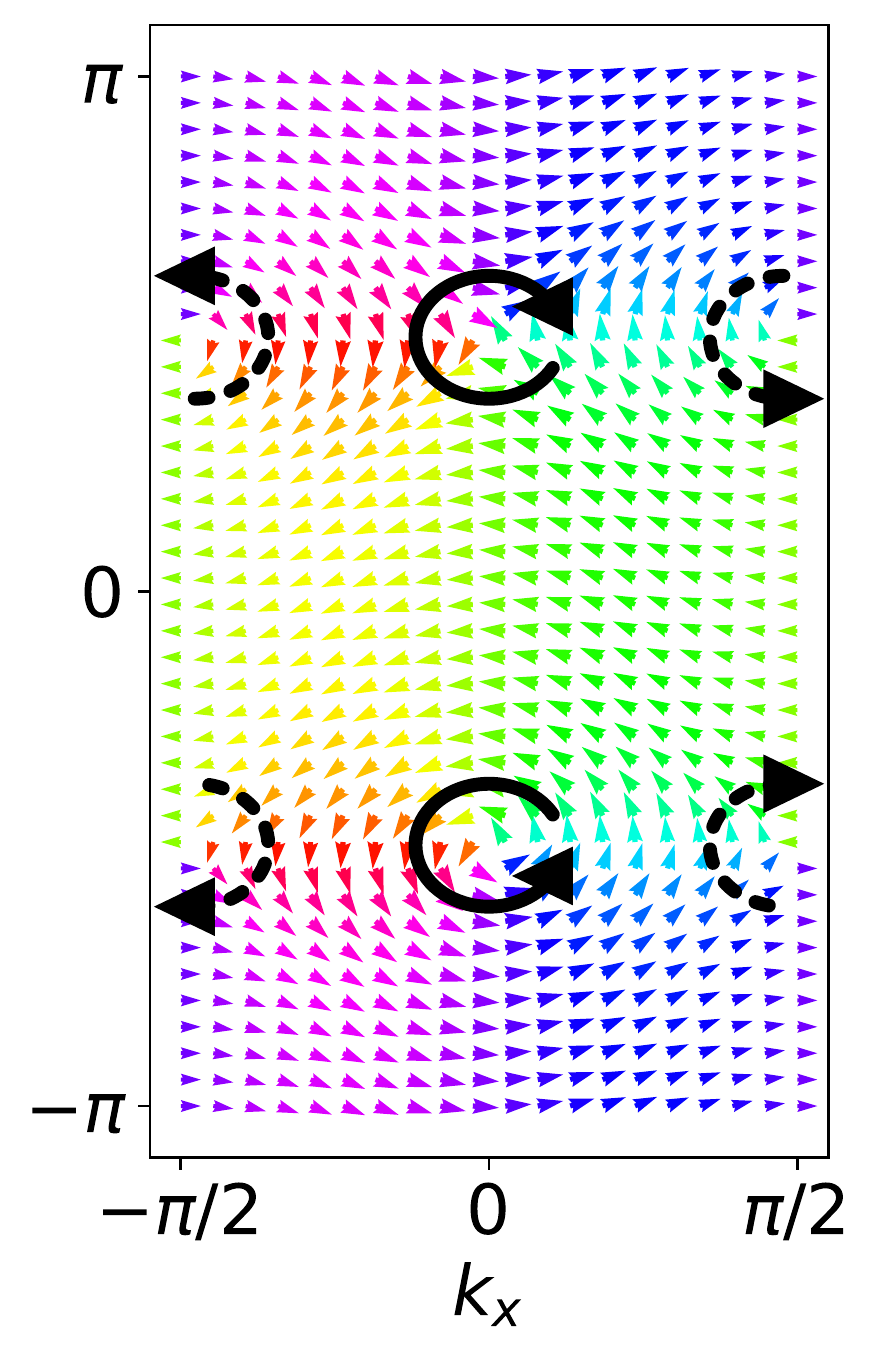}
  \label{subfig:q_2_rot_sub}
}
\caption{(a) Vortex field for $p/q=1/2$ of the upper band. (b) Vortex field of the lower band. (c) Vortex field of the rotated Hofstadter model for the upper band. (d) Vortex field of the rotated Hofstadter model for the lower band. Vortices are marked as oriented arrows, Dirac cones as dotted oriented arrows or as dotted circles if they lack a winding number. For the color code see Fig.~\ref{fig:QWZ}.}
\label{fig:q_2}
\end{figure*}

For $p/q = 1/2$ Eq.~(\ref{hofstadter_matrix}) becomes
\begin{equation}
h = \begin{pmatrix}\label{hofstadter_q_2}
2t_b \cos k_\text{y} & -t_a (1 + \e^{-2\ci k_\text{x}})\\
-t_a (1 + \e^{2\ci k_\text{x}}) & -2t_b \cos k_\text{y}
\end{pmatrix}.
\end{equation}
Two Dirac cones emerge at $(\pi/2, \pi/2)$ and $(\pi/2, -\pi/2)$, see Fig.~\ref{fig:E_2}. Their respective Hamiltonians in linear approximation are
\begin{equation}\label{linearized_q_2}
h_{\pm}(\vec{k}) = 2t_a \left(k_\text{x} - \frac{\pi}{2} \right) \sigma_\text{y} \pm 2t_b \left( k_\text{y} \pm \frac{\pi}{2} \right) \sigma_\text{z}.
\end{equation}
For two-band models only non-vanishing $f_\text{x}$ and $f_\text{y}$ around the Dirac point can lead to a definable winding number. Here, with $f_\text{x} = 0$ the introduction of winding vectors is expedient as we discuss below.

Close to the Dirac cone of an effective Hamiltonian $\hat{H} = \vec{f} \vec{\sigma}$ we can expand the Bloch vector $\vec{f}$ to linear order in polar coordinates
\begin{equation}
f_j(k_\text{x}, k_\text{y}) \approx \alpha_j k_\text{x} + \beta_j k_\text{y} = k \left( \alpha_j \cos \varphi + \beta_j \sin \varphi \right).
\end{equation}
Then for fixed $k = |\vec{k}|$ the Bloch vector traces the shape of an ellipse
\begin{equation}
\vec{f} = k (\vec{\alpha}\cos \varphi + \vec{\beta}\sin \varphi).
\end{equation}
Orthogonal to the plane spanned by $\vec{\alpha}$ and $\vec{\beta}$ we define the winding vector $\vec{w}$ as an oriented normal vector
\begin{equation}
\vec{w} = \frac{\vec{\alpha} \times \vec{\beta}}{|\vec{\alpha} \times \vec{\beta}|}.
\end{equation}
This definition is equivalent to the definition given in~\cite{mon}:
\begin{equation}
\vec{w} = \frac{1}{2\pi} \oint \vec{n} \times \diff \vec{n},
\end{equation}
where $\vec{n} = \vec{f} / |\vec{f}|$. For convenience, we work with a normalized vector $|\vec{\alpha} \times \vec{\beta}| = 1$. The phase $\phi$ of the second component of Eq.~(\ref{general_2_times_2}) ($f_\text{x} + \ci f_\text{y} = R \e^{\ci \phi}$) determines the winding number:
\begin{equation}
\phi = \arctan \left( \frac{f_\text{y}}{f_\text{x}} \right) = \arctan \left( \frac{\alpha_2 \cos \varphi + \beta_2 \sin \varphi}{\alpha_1 \cos \varphi + \beta_1 \sin \varphi} \right)
\end{equation}
and to have a well-defined winding orientation, $(\alpha_1, \beta_1)$ and $(\alpha_2, \beta_2)$ have to be linearly independent:
\begin{equation}\label{determinant_winding}
\det \begin{pmatrix}
\alpha_1 & \beta_1 \\
\alpha_2 & \beta_2
\end{pmatrix}= w_\text{z} \neq 0.
\end{equation}
The sign of $w_\text{z}$ in Eq.~(\ref{determinant_winding}) determines the winding orientation, analogously to the definition of chirality in~\cite{sim, kau}. $w_\text{z} > 0$ ($w_\text{z} < 0$) is related to counterclockwise (clockwise) winding. Alternatively we could say that the winding is $+1~(-1)$. For the Hamiltonian $\hat{H}_{M = 0} = k_\text{x} \sigma_\text{x} + k_\text{y} \sigma_\text{y}$ the winding vector would be $\vec{w} = \vec{e}_\text{z}$ and the winding is counterclockwise.

For the Dirac cones in Eq.~(\ref{linearized_q_2}) the winding vector is parallel to the $\sigma_\text{x}$-axis. To get a well-defined winding with respect to the $\sigma_\text{x}$–$\sigma_\text{y}$-plane we rotate the Hamiltonian around the $\sigma_\text{y}$-axis by $\pi/2$. Such a rotation corresponds to a unitary transformation that leaves the band spectrum invariant. Since a rotation affects the Hamiltonian like a continuous deformation, the Chern numbers of the bands must also be invariant. With the rotation matrices~\cite{zee}
\begin{equation}
\hat{R}_j(\omega) = \exp \left(\ci \frac{\omega}{2} \sigma_j \right) = \openone \cos \frac{\omega}{2} + \sigma_j \ci\sin \frac{\omega}{2}
\end{equation}
we find the transformed Hamiltonian
\begin{equation}
\begin{aligned}[b]
\hat{R}_\text{y}^{\dagger}(\pi/2) h_{\pm} \hat{R}^{\pdagger}_\text{y}(\pi/2) = &\pm 2t_b \left( k_\text{y} \pm \frac{\pi}{2} \right) \sigma_\text{x} \\
&+ 2 t_a \left( k_\text{x} - \frac{\pi}{2} \right) \sigma_\text{y}
\end{aligned}
\end{equation}
The Dirac cone at $k_\text{x} = k_\text{y} = \pi/2$ results in a vortex, the Dirac cone at $k_\text{x} = \pi/2, k_\text{y} = -\pi/2$ in an antivortex, see Fig.~\ref{subfig:q_2_rot_top} and \ref{subfig:q_2_rot_sub}. Their joint contribution to the Chern number is then zero. The unapproximated Hamiltonian rotated by $\pi/2$ is
\begin{multline}
  \hat{R}_\text{y}^{\dagger}(\pi/2) h \hat{R}^{\pdagger}_\text{y}(\pi/2) \\
  = \begin{pmatrix}
  t_a (1 + \cos 2k_\text{x}) & 2t_b \cos k_\text{y} + \ci t_a \sin 2k_\text{x} \\
  2t_b \cos k_\text{y} - \ci t_a \sin 2k_\text{x} & -t_a (1 + \cos 2k_\text{x})
  \end{pmatrix}.
\end{multline}
In the lower band there is now also a flux carrying anti\-vortex at $k_\text{x} = 0, k_\text{y} = \pi/2$ and a flux carrying vortex at $k_\text{x} = 0, k_\text{y} = -\pi/2$. This vortex–antivortex pair emerges with the rotation of the Hamiltonian and does not change the difference between vortices and antivortices. Adding all contributions together, the Chern number of the two bands is zero, as it is expected to be for any two-band system. Finally, note that in the unrotated vortex fields shown in Fig.~\ref{subfig:q_2_top} and \ref{subfig:q_2_sub} there appears to be a line at $k_\text{x} = \pi/2$ between $k_\text{y} = -\pi/2$ and $k_\text{y} = \pi /2$, where the vortex field vanishes. This can be associated to a $\mathbb{Z}_2$ invariant, as shown in Appendix~\ref{appendix_4}.

\subsection{Solution for arbitrary \boldmath{$p/q$}}\label{subsec:hofstadter_q_not_2}

The Hofstadter model has the advantage that the Chern number of every band for all possible $p, q$ can be computed analytically. This can even be achieved without using the Berry formalism. With the Streda formula one can calculate the Chern number of an isolated band~\cite{str}
\begin{equation}\label{chern_number}
C_{r} = t_r - t_{r-1},
\end{equation}
where $r$ denotes the $r$-th eigenspace, which are ordered according to their eigenvalues. If the band is gapped, then $r$ ($r-1$) may equivalently label the energy gap above (below) the band. $t_r$ can be determined with a Diophantine equation
\begin{equation}\label{diophantine}
r = q s_r + p t_r,
\end{equation}
where $s_r, t_r \in \mathbb{Z}$ and $|t_r| \leq q/2$. It is also possible to prove Eq.~(\ref{chern_number}) using the Berry formalism, see Ref.~\cite{tknn, fra, ber, koh2}. However, there the two bands connected by Dirac cones have been ignored. We want to address this deficit.

\begin{figure}
  \centering
  \includegraphics[width=1.0\linewidth]{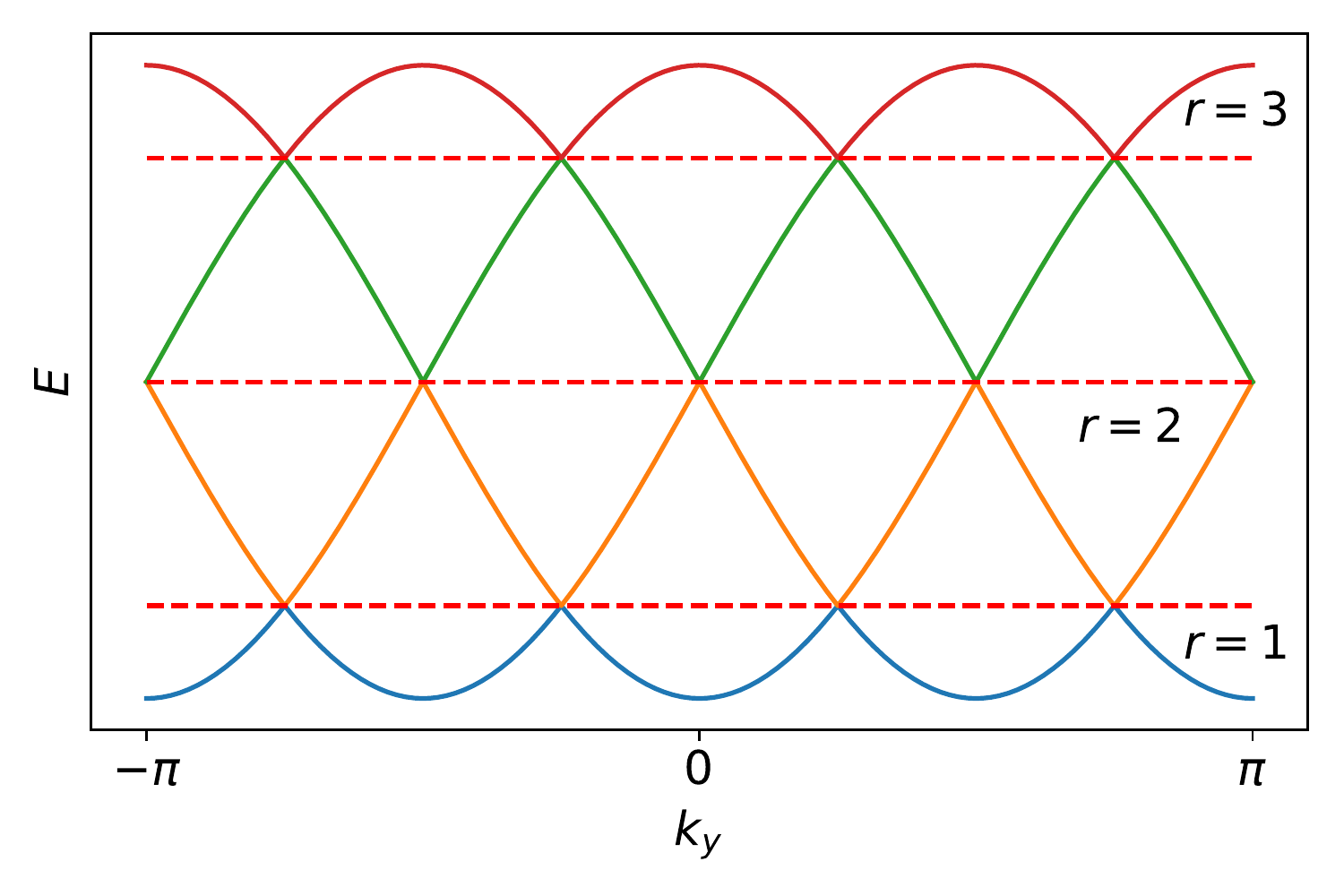}
  \caption{Energy spectrum of the Hofstadter model for $q=4$, $t_a = 0$.}\label{fig:bands_with_gaps}
\end{figure}

Chern numbers can only change as band gaps close and reopen. We use this to give the Hamiltonian a simpler form. In the limit $t_a \to 0$, $h_{\mu, \mu'}$ becomes diagonal and the eigenenergies become cosine bands with no dispersion along the $k_\text{x}$-axis, see Fig.~\ref{fig:bands_with_gaps}. To these bands we associate eigenstates $c_{\mu}^{\dagger} \ket{0} = \ket{\mu}$ with eigenenergies $E_{\mu}(k_\text{y}) = -2t_b \cos(k_\text{y} + 2\pi \varphi \mu)$. Note that these states are not the Bloch bands, which are ordered according to their eigenvalues; to distinguish the two, we assign the Bloch bands a latin index, whereas the eigenstates with the continuous cosine dispersion will be denoted using greek indices. In the weak coupling regime $t_a \ll t_b$ eigenstates $\ket{m}$ should approximately be equal to $\ket{\mu}$ for some $\mu$, except in the neighborhood of a degeneracy between two bands $\mu_1$ and $\mu_2$. Here, eigenstates are obtained by an effective $2 \times 2$ matrix $h_{r, \alpha}$, with $(h_{r, \alpha})_{j,l} = \braket{\mu_j|\hat{h}_{k_\text{x}, k_\text{y}}|\mu_l}$. The index $\alpha$ runs from $1$ to $q$, thereby covering all $q$ crossings of a gap $r$, see Fig.~\ref{fig:bands_with_gaps}. The Bloch states $\ket{m}$ have to be gained in perturbation theory and $h_{r, \alpha}$ has off-diagonal elements that cause hybridization. Details are given in Appendix~\ref{appendix_3}. In the vicinity of the $r$-th gap opening, there are $|t_r|$ matrices of the form
\begin{equation}\label{effective_1}
h_{r, \alpha} = \begin{pmatrix}
\epsilon & \varDelta_r \e^{-\text{sgn}(t_r) \ci k_\text{x} q} \\
\varDelta_r \e^{\text{sgn}(t_r) \ci k_\text{x} q} & -\epsilon
\end{pmatrix}
\end{equation}
and $q - |t_r|$ matrices of the form
\begin{equation}\label{effective_trivial_1}
h_{r, \alpha} = \begin{pmatrix}
\epsilon & \varDelta_r \\
\varDelta_r & -\epsilon
\end{pmatrix},
\end{equation}
where $\varDelta_r$ is a small hybridization matrix element that scales like $t_a^{|t_r|}/t_b^{|t_r|-1}$ and $\epsilon = v_\text{y} k_\text{y}$ with $v_\text{y} > 0$, see Eq.~(\ref{m_1_m_2_element}). For convenience, we omit the explicit dependence of $\epsilon$ on $r$ and $\alpha$.

\begin{figure}
  \centering
  \includegraphics[width=1.0\linewidth]{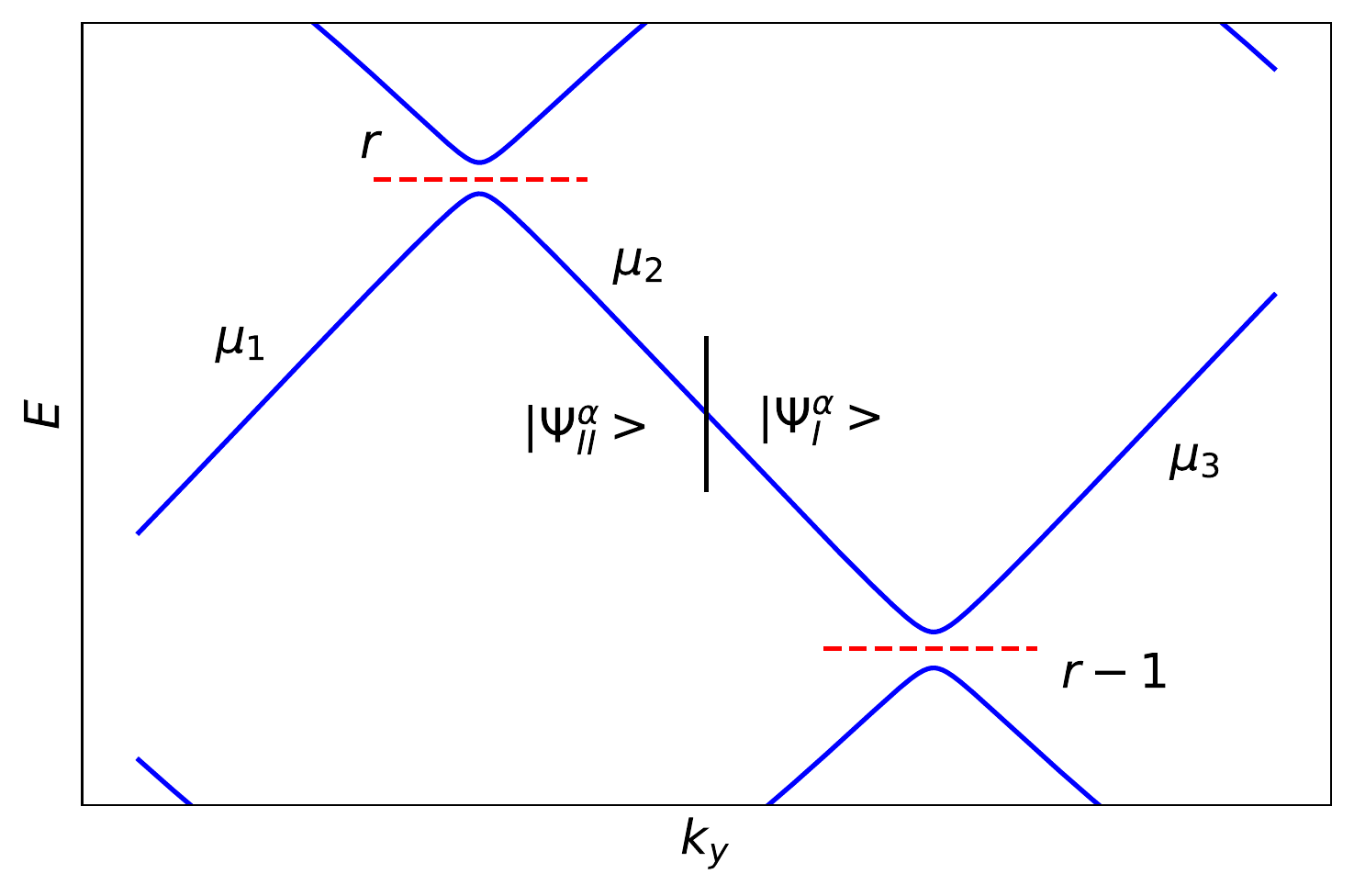}
  \caption{Closeup of the phase convention chosen between the $r$-th and $r-1$-th gap. The vertical bar separates two neighboring patches which have to be defined for intervals of $2\pi/q$ in $k_{\text{y}}$-direction. $\ket{\Psi_{\text{I}}^{\alpha}}$ and $\ket{\Psi_{\text{II}}^{\alpha}}$ are two eigenstates of the same band, but with a different phase convention. $\ket{\Psi_{\text{II}}^{\alpha}}$ has a positive real $\ket{\mu_1}$ component and $\ket{\Psi_{\text{I}}^{\alpha}}$ a positive real $\ket{\mu_3}$ component.}\label{fig:bands_zoom_patch}
\end{figure}

For gapped bands the vortex field frame is not yet required and we determine the Chern number by partitioning the Brillouin zone into patches as described in section~\ref{subsec:Gapped_bands}. Around each degeneracy we choose the phase convention to make the first component with the strictly increasing energy dispersion positive real. We need the negative energy eigenstates for the band below the $r$-th gap. The eigenstates of Eq.~(\ref{effective_1}) are then given as $\ket{\Psi_{\text{II}}^{\alpha}} = \ket{\mu_1} + 0^{+} \ket{\mu_2}$ for $\epsilon < 0$ and $\ket{\Psi_{\text{II}}^{\alpha}} = 0^{+} \ket{\mu_1}  -\text{sgn}(\varDelta_r) \e^{\text{sgn}(t_r) \ci k_\text{x} q} \ket{\mu_2}$ for $\epsilon > 0$, see Fig.~\ref{fig:bands_zoom_patch}. ``$0^{+}$'' symbolizes the fact that we work with eigenstates far enough away from the degeneracy that one of the components can be made arbitrarily small, yet it never goes exactly to zero. Correspondingly, the eigenstates for the $(r-1)$-th gap are $\ket{\Psi_{\text{I}}^{\alpha}} = 0^{+} \ket{\mu_3} + \text{sgn}(\varDelta_{r-1}) \e^{\text{sgn}(t_{r-1}) \ci k_\text{x} q} \ket{\mu_2}$ for $\epsilon < 0$ and $\ket{\Psi_{\text{I}}^{\alpha}}= \ket{\mu_3} + 0^{+} \ket{\mu_2}$ for $\epsilon > 0$, as here we have to consider the positive energy eigenstates. The eigenstates of Eq.~(\ref{effective_trivial_1}) do not carry relevant phase factors~\footnote{Note that technically to conform with our perturbation ansatz we need one further patch line the decreasing sections of the band, as we have made only the parts with increasing dispersion real. However, it is easy to convince oneself that this will not change the end result, as it is ultimately equivalent to have one transition function between $\ket{\Psi_{\text{II}}^{\alpha}}$ and $\ket{\Psi_{\text{I}}^{\alpha}}$ or to have two transition functions that connect the states with a state, where the decreasing component is non-negative real and analogously for eigenstates of Eq.~(\ref{effective_trivial_1}).}.

We need $q$ patches, each one covering both a degeneracy of the $(r-1)$-th gap at some fixed $k_\text{y}$ and a degeneracy of the $r$-th gap at $k_\text{y} + \pi/q$.
Eq.~(\ref{Chern_number_winding}) becomes
\begin{equation}\label{Chern_number_patches_Hofstadter}
C_{r} = \frac{1}{2\pi} \displaystyle\sum_{\alpha=1}^{q} \int_{\pi/q}^{-\pi/q} \diff k_\text{x}  \frac{\diff \left( \Arg(\e^{\ci\chi_{\alpha}}) \right)}{\diff k_\text{x}}.
\end{equation}
We define $\e^{\ci \chi_{\alpha}} = \e^{\ci \chi_{r-1, \alpha}} \e^{-\ci \chi_{r, \alpha}}$, with $\ket{\Psi_{\text{I}}^{\alpha}} = \e^{\ci \chi_{\alpha}} \ket{\Psi_{\text{II}}^{\alpha}}$. In $|t_r|$ cases $\e^{\ci \chi_{r, \alpha}} = - \text{sgn}(\varDelta_r) \exp(\text{sgn}(t_r) \ci k_\text{x} q)$ and in $|t_{r-1}|$ cases $\e^{\ci \chi_{r-1, \alpha}} = \text{sgn}(\varDelta_{r-1}) \exp(\text{sgn}(t_{r-1}) \ci k_\text{x} q)$. In all other cases, where the crossing is of the form of Eq.~(\ref{effective_trivial_1}), $\e^{\ci \chi_{\alpha}}$ is equal to $\pm 1$. The Chern number is then equal to $C_{r} = t_r - t_{r-1}$, like in Eq.~(\ref{chern_number}).

\begin{figure*}[tb]
\subfloat[]{
  \includegraphics[width=0.95\textwidth]{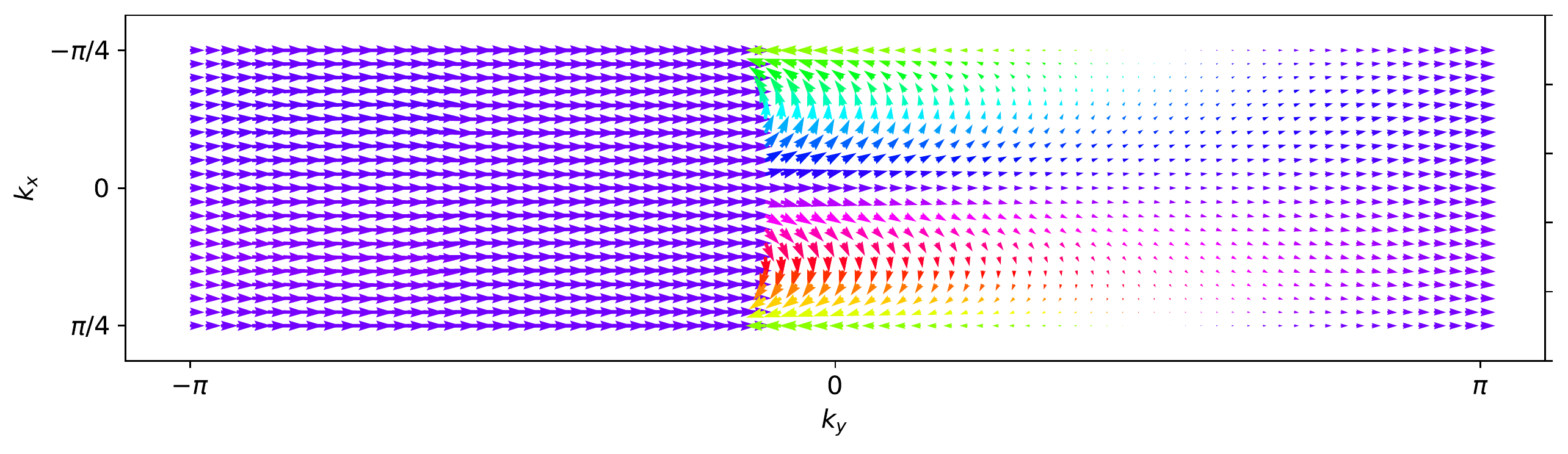}
 \label{fig:patches_q_4}
}\hfil
\subfloat[]{
  \includegraphics[width=0.95\textwidth]{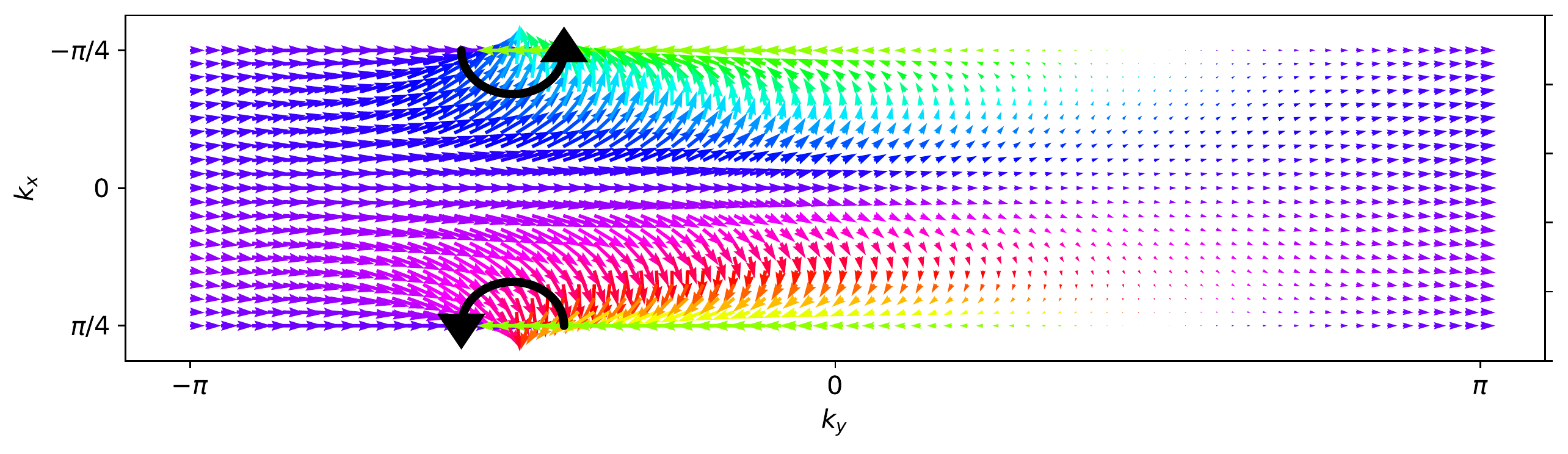}
 \label{fig:vortex_q_4}
}
\caption{Patch construction vs. vortex field for the $p/q=1/4$ Hofstadter model. (a) Plot of the first component of the eigenvectors of the lowest band. On each patch the phase convention is chosen as described in the main text. To make the phase difference between neighboring patches easier to spot, the component is plotted for $t_a = t_b = 1$, which is qualitatively the same as in the weak coupling limit. The boundaries of the patches are located at $k_\text{y} = -5\pi/8, -\pi/8, 3\pi/8, 7\pi/8$. The patches touching at $k_\text{y} = -\pi/8$ are connected in the weak coupling limit with the transition function $\e^{\ci \chi_{\alpha}} = \e^{-\ci k_\text{x} q}$ which leads to the correct Chern number $C_{1} = 1 = t_{1} - t_{0}$ (with $t_0 = t_q = 0$) according to Eq.~(\ref{Chern_number_patches_Hofstadter}). The other transition functions in the limit $t_a \to 0$ are equal to one. (b) Vortex field for the lowest band with $\ket{\Phi_{\text{II}}} = \ket{1}$ and $\ket{\Phi_{\text{I}}} = \ket{3}$. The vortex at $k_\text{x} = \pi/4$, $k_\text{y} = - \pi/2$ corresponds to the correct Chern number $C_{1} = 1$. For the color code see Fig.~\ref{fig:QWZ}.}
\end{figure*}

An example is plotted in Fig.~\ref{fig:patches_q_4} for $p/q=1/4$ and the lowest band. It is easy to spot the phase difference between the gauges at $k_\text{y} = -\pi/8$, which leads to the correct Chern number of $C_{1} = 1$, because the only non-trivial $\e^{\ci \chi_{\alpha}}$ is equal to $\e^{-\ci k_\text{x} q}$ in the limit $t_a \ll t_b$. Alternatively one could have considered a vortex field, which is plotted in Fig.~\ref{fig:vortex_q_4}.

If $q$ is an even number, then at $E = 0$ there are $q$ Dirac cones~\cite{wen} that also manifest themselves in the band hybridization elements. If $q = 4n$, $n \in \mathbb{N}$ then at $r = q/2$ they are
\begin{equation}\label{dirac_sine}
h_{q/2, \alpha_{\pm}} =
\begin{pmatrix}
\epsilon & \pm \ci \varDelta_{\frac{q}{2}} \e^{\mp \ci k_\text{x} q/2} \sin (k_\text{x} q/2) \\
\text{c.c.} & - \epsilon
\end{pmatrix},
\end{equation}
where $\alpha_{+} = \alpha_{-} + q/2$, see Eq.~(\ref{hybridization_sin}). In other words, for every degeneracy at $k_\text{y}$ there is another degeneracy at $k_\text{y} + \pi$ of the same subspace but with a complex conjugated coupling matrix. If $q = 2(2n + 1)$, then
\begin{equation}\label{dirac_cosine}
h_{q/2, \alpha_{\pm}} = \begin{pmatrix}
\epsilon &  \varDelta_{\frac{q}{2}} \e^{\mp \ci k_\text{x} q/2} \cos (k_\text{x} q/2) \\
\text{c.c.} & - \epsilon
\end{pmatrix},
\end{equation}
see Eq.~(\ref{hybridization_cos}). The winding vectors of the Dirac cones again point toward the $\sigma_\text{x}$ axis of each subspace. Therefore, we rotate all subspaces around their $\sigma_\text{y}$ axes by $\pi/2$. Since all the rotation matrices $\hat{R}_{\text{y}, \mu_1}(\pi/2)$ for different subspaces $\lbrace \ket{\mu_1}, \ket{\mu_2} = \ket{\mu_1+q/2} \rbrace$ commute with each other, there is no particular order and we can build the product of all of them:
\begin{equation}
\hat{R}_{\text{tot}} = \displaystyle\prod_{\mu = 1}^{q/2} \hat{R}_{\text{y}, \mu_{1}}(\pi/2) = \frac{1}{\sqrt{2}} \left( \begin{array}{c|c}
\openone & \openone \\
\hline
-\openone & \openone\\
\end{array} \right).
\end{equation}
$\openone$ are $q/2 \times q/2$ unit matrices. Notice that the winding numbers of the two Dirac cones in the same subspace are opposite to each other as in the case of $p/q=1/2$.

We want to calculate the Chern number $C$ of the two bands with Dirac cones. We will have the contributions of the gaps with $r = q/2-1$ and $r = q/2 + 1$ that have to be treated as we did before. Additionally, we have the contributions of the Dirac cones. The optimal route is therefore to split up the patches we worked with before. There used to be $q$ patches for a given band, covering a degeneracy at $r$ and $r-1$; now we will work with $2q$ patches, one for each degeneracy.

What is the transition function $\e^{\ci \chi_{\alpha}}$ between the patches? Without loss of generality we start with the lower band $m = q/2$. Let us denote with $\ket{\Psi_{0}^{q/2}}$ the eigenstates of the Hofstadter matrix $h$ (see Eq.~(\ref{hofstadter_matrix})) and with $\ket{\Psi_{\pi/2}^{q/2}}$ the eigenstates of $\hat{R}_{\text{tot}}^{\dagger} h \hat{R}^{\pdagger}_{\text{tot}}$. Both eigenstates are connected like $\ket{\Psi_{\pi/2}^{q/2}} = \hat{R}_{\text{tot}}^{\dagger} \ket{\Psi_{0}^{q/2}}$. The transition function is then defined by
\begin{equation}
\begin{aligned}[b]
\Arg(\e^{\ci \chi}) &= \Arg\left( \braket{\Psi_{\text{II}, \pi/2}^{q/2}|\Psi_{\text{I}, \pi/2}^{q/2}} \right) \\
&= \Arg\left( \braket{\Phi_{\text{II}}|\hat{P}^{q/2}_{\Psi_{\pi/2}}|\Phi_{\text{I}}} \right),
\end{aligned}
\end{equation}
with the projector $\hat{P}^{q/2}_{\Psi_{\pi/2}}$ onto the state $\ket{\Psi_{\pi/2}^{q/2}}$. By substituting $\ket{\Phi_{\text{II}}} \to \hat{R}_{\text{tot}}^{\dagger} \ket{\Phi_{\text{II}}}$ and using the idempotence of the projector we find
\begin{equation}
\Arg(\e^{\ci\chi}) = \Arg\left( \braket{\Phi_{\text{II}}|\Psi^{q/2}_{0}} \braket{\Psi_{0}^{q/2}|\hat{R}_{\text{tot}}|\Psi_{\pi/2}^{q/2}} \braket{\Psi_{\pi/2}^{q/2}|\Phi_{\text{I}}} \right).
\end{equation}
We have determined earlier that the only form of $\ket{\Psi_{0}^{q/2}} \braket{\Psi_{0}^{q/2}|\Phi_{\text{II}}}$ that results in non-trivial contributions is $0^{+} \ket{\mu - t_{q/2 - 1}} + \text{sgn}(\varDelta_{q/2-1}) \e^{\text{sgn}(t_{q/2-1}) \ci k_\text{x} q} \ket{\mu}$. For $\hat{R}_{\text{rot}} \ket{\Psi_{\pi/2}^{q/2}} \braket{\Psi_{\pi/2}^{q/2}|\Phi_{\text{I}}}$ one finds a solution of the form $a\ket{\mu} + b\ket{\mu + q/2}$. It is easy to show, that either $\Re (a) < 0$ or $>0$ and similarly for $b$. Therefore,
\begin{equation}
\begin{aligned}[b]
\diff \Arg(\e^{\ci \chi}) &= \diff \Arg\left(\braket{\mu| \text{sgn}(\varDelta_{q/2 - 1}) \e^{- \text{sgn}(t_{q/2 - 1}) \ci k_\text{x} q} a|\mu}\right) \\
&= \text{sgn}(t_{q/2-1}) q \diff k_\text{x} + \diff \Arg(a),
\end{aligned}
\end{equation}
where $\int_{\pi/q}^{-\pi/q} \diff k_{\text{x}} \partial\Arg(a)/\partial k_{\text{x}} = 0$. The upper band with $m = q/2 + 1$ can be treated analogously. Therefore, the contributions from the patches of the lower and upper band combined are the same as in the gapped case, $t_{q/2 + 1} - t_{q/2 - 1}$.

We still have to analyze the patches containing the Dirac cones. We make use of a vortex field given by
\begin{equation}
\begin{aligned}[b]
\Arg(\e^{\ci \chi}) &= \Arg\left(\braket{\mu + q/2|\hat{P}^{q/2}_{\Psi_{\pi/2}}|\mu} \right)\\
&= \Arg(\pm \epsilon -\ci \varDelta/2 \sin(k_\text{x} q)).
\end{aligned}
\end{equation}
It results in two vortices with opposite winding in every patch. One of the vortices is pinned by the Dirac point, the other one is shifted by $\pi/2$ in $k_\text{x}$-direction, where the band is gapped. This can be seen for the $p/q=1/2$ case in Fig.~\ref{subfig:q_2_rot_top},~\ref{subfig:q_2_rot_sub}. Independent of the way we split the Dirac cones up with a mass term, there must be exactly one clockwise and one counterclockwise flux carrying vortex in one of the two bands at the position of the former Dirac cones. Therefore, the net contribution of both vortices at the Dirac cones is zero and we are left with the contributions from the gaps at $q/2-1$ and $q/2 + 1$. The total Chern number is then $C = t_{q/2+1} - t_{q/2 - 1}$.

\section{Conclusion}\label{sec:Conclusion}

In this paper we have studied the effect of topological defects on Chern numbers of gapless bands. These defects manifest themselves as Dirac cones that can be characterized with winding vectors. This generalizes the notion of winding numbers, which cannot be defined if the winding vector is placed in the $\sigma_\text{x}$–$\sigma_\text{y}$-plane. Such a situation occurs in the Hofstadter model. We have shown that in a weak coupling limit all of its Dirac cones lack a well-defined winding number. In that case Chern numbers of bands can be calculated by rotating their characteristic winding vectors $\vec{w}$ to have a non-vanishing $\sigma_\text{z}$-component. Then a combination of single-band vortex fields of the involved bands leads to the correct Chern number. This procedure has been successfully applied on the Hofstadter model which to our knowledge has not been done in the literature.

The next logical step for future research would be to extend the notion of winding vectors and winding numbers to Dirac cones and topological defects in Hermitian $q \times q$, $q > 2$ matrices. It is possible to work with effective $2 \times 2$ matrices by projecting the Hamiltonian onto the eigenbasis of the contact point of the Dirac cone, as outlined in~\cite{mon, kau}: With the contact point $\vec{k}_0$ the Hamiltonian can be written as $\hat{H} = \hat{H}(\vec{k}_0) + \hat{H}_1(\vec{k} - \vec{k}_{0}) + \mathcal{O}((\vec{k} - \vec{k}_{0})^2)$. If $U(\vec{k}_0)$ diagonalizes $\hat{H}(\vec{k}_0)$ and $\hat{P}$ projects onto the eigenbasis of the contact point of $\hat{H}(\vec{k}_{0})$ then $\hat{P}U^{\dagger}(\vec{k}_0) \hat{H}_1(\vec{k} - \vec{k}_{0})U(\vec{k}_0)\hat{P}$ provides an effective $2\times 2$ Hamiltonian of a Dirac cone that can be attributed a winding vector. A promising study ground for this ansatz could again be the Hof\-stadter model. We have already found earlier that in the limit $t_a \to 0$ two Dirac cones located at $k_\text{y}$ and $k_\text{y} + \pi$ exist in the same eigenspace and have opposite winding. We suspect that this might be the case in the isotropic case, too, as one finds
\begin{equation}
\hat{H}_{1}(k_\text{x} - k_\text{x}^{0}, k_\text{y} - k_\text{y}^{0}) = \hat{H}_{1}(k_\text{x} - k_\text{x}^{0}, -k_\text{y} - (k_\text{y}^{0} + \pi))
\end{equation}
and
\begin{equation}
U(k_{\text{x}}^{0}, k_\text{y}^{0} + \pi) = \left( \begin{array}{c|c}
0 & \openone \\
\hline
(-1)^{q/2} \openone & 0\\
\end{array} \right) U(\vec{k}_{0}).
\end{equation}
This study may be interesting for inquiring merging scenarios of Dirac cones beyond $p/q = 1/2$~\cite{del}.

Finally, it should be noted that the vortex field approach does not provide us with a new type of topological invariant but a more efficient way to compute Chern numbers that shows its strength if the position of vortices can be determined based on some symmetries. It would be interesting to use this fact in computationally expensive problems like in the study of correlated Chern insulators, where symmetries can indeed be used to predict the location of a topological phase transition in an associated manifold~\cite{var}.

\begin{acknowledgments}

We wish to thank D.~Braak and T.~Weth for stimulating discussions. This work was supported by the Deutsche Forschungsgemeinschaft (DFG) through TRR80 (project number 107745057), and through QUAST-FOR5249 - 449872909 (project TP4).

\end{acknowledgments}

\appendix

\section{Magnetic translation operators}\label{appendix_1}

The Hamiltonian of electrons subject to a periodic potential has a common set of eigenstates with the translation operators $\hat{T}_{\vec{a}} = \e^{-\ci \vec{a} \hat{\vec{p}}/\hbar}$ if $\vec{a}$ is a unit vector of the lattice. In the presence of an external magnetic field this does not hold anymore, because the vector potential $\vec{A}(\vec{r})$ is aperiodic. However, the vector potential on two lattice sites can be related by a gauge transformation $\phi_{\vec{a}}(\vec{r})$:
\begin{equation}
\vec{A}(\vec{r} - \vec{a}) = \vec{A}(\vec{r}) + \vec{\nabla} \phi_{\vec{a}}(\vec{r})
\end{equation}
Thereby, we can define magnetic translation operators that commute with the Hamiltonian~\cite{fra, zak}
\begin{equation}
\hat{T}_{\vec{a}}^{\text{M}} = \exp\left( \ci \frac{e}{\hbar c} \phi_{\vec{a}}(\vec{r}) \right) \hat{T}_{\vec{a}}.
\end{equation}
$\hat{T}_{\vec{a}}^{\text{M}}$ and $\hat{T}_{\vec{b}}^{\text{M}}$ commute only if $\vec{a}$ and $\vec{b}$ span an area threaded by an integer number of flux quanta, so for $p/q$ flux quanta per lattice site the (magnetic) unit cell of the Hofstadter Hamiltonian has to contain $q$ sites for all involved magnetic translation operators to be diagonalizable simultaneously with the Hamiltonian. In the gauge convention of Eq.~(\ref{real_space_hofstadter}) we identify $\hat{T}_{\vec{e}_\text{y}}^{\text{M}} = \hat{T}_{\vec{e}_{\text{y}}}$ and
\begin{equation}\label{T_M_x_relation}
\begin{aligned}[b]
\hat{T}_{\vec{e}_\text{x}}^{\text{M}} &= \displaystyle\sum_{m,n,\mu} \e^{-2\pi \ci n \varphi} \left( c_{m,n}^{\mu+1} \right)^{\dagger} c_{m,n}^{\mu} \\
&= \displaystyle\sum_{k_\text{x}, k_\text{y}, \mu} \left( c_{k_\text{x}, k_\text{y} - 2\pi \varphi}^{\mu + 1} \right)^{\dagger} c_{k_\text{x}, k_\text{y}}^{\mu}.
\end{aligned}
\end{equation}
As unit cell vectors we choose $\vec{e}_{\text{y}}$ and $q \vec{e}_{\text{x}}$, because then $\hat{T}^{\text{M}}_{q \vec{e}_{\text{x}}} = \hat{T}_{q \vec{e}_\text{x}}$. This implies $2\pi/q$ periodicity of the energy spectrum in $k_{\text{x}}$ direction. Translating an electron around a unit cell results in an Aharonov–Bohm phase
\begin{equation}
\hat{T}_{\vec{e}_\text{x}}^{\text{M}} \hat{T}_{\vec{e}_\text{y}}^{\text{M}} \left( \hat{T}_{\vec{e}_\text{x}}^{\text{M}}\right)^{-1} \left( \hat{T}_{\vec{e}_\text{y}}^{\text{M}}\right)^{-1} = \e^{-2\pi \ci \varphi}.
\end{equation}
As a consequence, if $\ket{m, k_\text{x}, k_\text{y}}$ is a Bloch eigenstate of $\hat{H}$, then
\begin{equation}
\hat{T}^{\text{M}}_{\vec{e}_\text{y}} \hat{T}^{\text{M}}_{\vec{e}_\text{x}} \ket{m, k_\text{x}, k_\text{y}} = \e^{-\ci (k_\text{y} - 2\pi \varphi)} \hat{T}^{\text{M}}_{\vec{e}_\text{x}} \ket{m, k_\text{x}, k_\text{y}}
\end{equation}
which implies
\begin{equation}
\hat{T}^{\text{M}}_{\vec{e}_{\text{x}}} \ket{m, k_\text{x}, k_\text{y}} \propto \ket{m, k_\text{x}, k_\text{y} - 2\pi \varphi},
\end{equation}
so the energy spectrum is also $2\pi/q$-periodic in $k_\text{y}$-direction. We can characterize this state further. If
\begin{equation}
\ket{m, k_\text{x}, k_\text{y}} = \displaystyle\sum_{\mu = 1}^{q} b^{m}_{\mu} \left( c_{k_\text{x}, k_\text{y}}^{\mu} \right)^{\dagger} \ket{0},
\end{equation}
where $b^{m}_{\mu}(\vec{k})$ are the eigenvector components, then with Eq.~(\ref{T_M_x_relation})
\begin{equation}\label{t^m_commute}
\hat{T}_{\vec{e}_\text{x}}^{\text{M}} \ket{m, k_\text{x}, k_\text{y}} = \displaystyle\sum_{\mu = 1}^{q} b^{m}_{\mu - 1} \e^{-\ci k_\text{x} q \delta_{1, \mu}} \left( c_{k_\text{x}, k_\text{y} - 2\pi \varphi}^{\mu} \right)^{\dagger} \ket{0}
\end{equation}
with $b^{m}_{\mu} = b^{m}_{\mu + q}$ and the Kronecker delta $\delta_{1, \mu}$. This means that all components of the eigenvector permute if the eigenvector is shifted from $k_\text{y}$ to $k_\text{y} - 2\pi \varphi$; the $\mu = 1$ component is supplied with an additional phase factor to satisfy $c_{\mu + q}^{\dagger} = \e^{-\ci k_\text{x} q} c_{\mu}^{\dagger}$.

\section{Effective coupling order around unperturbed band degenarcies}\label{appendix_2}

In the weak coupling limit we write Eq.~(\ref{h_kx_ky}) as $\hat{h}_{k_\text{x}, k_\text{y}} = \hat{h}_{0, k_\text{x}, k_\text{y}} - t_a \hat{V}_{k_\text{x}, k_\text{y}}$ with
\begin{equation}
\begin{gathered}
\hat{h}_{0, k_\text{x}, k_\text{y}} = - 2 t_b \displaystyle\sum_{\mu = 1}^{q} \cos(k_\text{y} + 2\pi \varphi \mu) c_{\mu}^{\dagger} c^{\pdagger}_{\mu} \\
\hat{V}_{k_\text{x}, k_\text{y}} = \displaystyle\sum_{\mu = 1}^{q} \left( c_{\mu + 1}^{\dagger} c^{\pdagger}_{\mu} + c_{\mu}^{\dagger} c^{\pdagger}_{\mu + 1} \right).
\end{gathered}
\end{equation}
Because $\hat{V}_{k_\text{x}, k_\text{y}}$ only contains nearest neighbor hopping terms, to couple two bands $\ket{\mu_1}$ and $\ket{\mu_2}$ around a degeneracy at the $r$-th gap with $|\tilde{t}_r| := \text{min}(|\mu_2 - \mu_1|, q - |\mu_2 - \mu_1|)$ we need perturbation theory of $|\tilde{t}_r|$-th order. For now, we distinguish $\tilde{t}_{r}$ and $t_r$ from the Diophantine equation in Eq.~\ref{diophantine}, because we have not shown the equivalence yet. There is a connection between the algebra of magnetic translation operators and the Diophantine equation~\cite{dan}. We inquire
\begin{equation}\label{t_r_derivation}
\begin{aligned}[b]
 E_{\mu_1}(k_{\text{y}}) &= \braket{\mu_1, k_\text{x}, k_\text{y}|\hat{h}_{0} \left( \displaystyle\sum_{\mu = 1}^{q} c_{\mu + 1}^{\dagger} c^{\pdagger}_{\mu} \right)^{\tilde{t}_r}|\mu_2, k_\text{x}, k_\text{y}}\\
&= \braket{\mu_1, k_\text{x}, k_\text{y}|\hat{h}_{0} \left( \hat{T}_{\vec{e}_\text{x}}^{\text{M}} \right)^{\tilde{t}_r}|\mu_2, k_\text{x}, k_\text{y} + 2\pi \alpha/q},
\end{aligned}
\end{equation}
see Eq.~(\ref{t^m_commute}). Note that $\hat{T}_{\vec{e}_\text{x}}^{\text{M}}$ shifts $k_\text{y}$, requiring the addition of the term $2\pi \alpha/q$ in $\ket{\mu_2, k_\text{x}, k_\text{y} + 2\pi \alpha/q}$. The eigenenergy, where the $r$-th gap opens, is $E_{\text{gap}} = -2t_b \cos(r \pi/q)$. Then, if $k_\text{y}$ is the wavevector at the degeneracy, where we find $E_{\mu_1}(k_{\text{y}}) = E_{\mu_2}(k_{\text{y}})$, according to Eq.~(\ref{t_r_derivation}) (because $\hat{T}^{\text{M}}_{\vec{e}_{\text{x}}}$ commutes with $\hat{h}_{0}$) $E_{\text{gap}} = E_{\mu_2}(k_\text{y}) = E_{\mu_2}(k_\text{y} + 2\pi \alpha / q)$:
\begin{equation}
\cos(r \pi/q) = \cos(k_\text{y} + 2\pi \varphi \mu_2) = \cos(k_\text{y} + 2\pi \alpha/q + 2\pi \varphi \mu_2)
\end{equation}
and therefore
\begin{equation}\label{relation_alpha_r}
r = (\pm)_{k_\text{y}, \mu_2} \alpha ~(\text{mod }q),
\end{equation}
where
\begin{equation}\label{plus_minus_k_y_m_2}
(\pm)_{k_\text{y}, \mu_2} =  \begin{cases}
+ &\text{if } k_\text{y} + 2\pi \varphi \mu_2 ~ (\text{mod }2\pi) \in (-\pi, 0)\\
- &\text{if } k_\text{y} + 2\pi \varphi \mu_2 ~ (\text{mod }2\pi) \in (0, \pi)\\
\end{cases}.
\end{equation}
We can deduce from Eq.~(\ref{t_r_derivation}) together with Eq.~(\ref{T_M_x_relation}) a relation between $\tilde{t}_r$ and $\alpha$
\begin{equation}
k_\text{y} + \frac{2\pi \alpha}{q} - 2\pi \varphi \tilde{t}_{r} = k_\text{y} ~(\text{mod }2\pi).
\end{equation}
Together with Eq.~(\ref{relation_alpha_r}) this yields
\begin{equation}
\quad (\pm)_{k_\text{y}, \mu_2} r = p \tilde{t}_r + q s_r,
\end{equation}
where $s_r \in \mathbb{Z}$. Note the additional sign compared to Eq.~(\ref{diophantine}), hence $t_r = (\pm)_{k_\text{y}, \mu_2} \tilde{t}_r$. If $E_{\mu_2}(k_\text{y})$ is decreasing and $E_{\mu_1}(k_\text{y})$ is increasing at the degeneracy, then, according to Eq.~(\ref{plus_minus_k_y_m_2}), $(\pm)_{k_\text{y}, \mu_2} = +$ and $t_r = \tilde{t}_r$.

\section{Hofstadter model in the weak coupling limit}\label{appendix_3}

We want to deduce Eq.~(\ref{effective_1}) and Eq.~(\ref{effective_trivial_1}). To get well-defined matrix elements we demand that $E_{\mu_1}(k_\text{y})$ is monotonically increasing at the degeneracy, so $\tilde{t}_{r} = t_r$ and $\braket{\mu_1|\hat{h}_{k_\text{x}, k_\text{y}}|\mu_1} = v_\text{y} k_\text{y}$, with $v_\text{y} > 0$. For convenience, in the following we change notation and use $\ket{\mu}$ for the expanded states~\footnote{This will not cause issues as the dominant contribution to the expanded states comes from the original one in the limit $t_a \to 0$. Because of this the Berry curvature of the expanded states goes to zero in this limit, so the relevant topological information of the Bloch states is contained in the matrices of Eq.~(\ref{effective_1}), not in their basis.}. Unperturbed states and energies will now contain a superscript $(0)$, i.e. $\ket{\mu^{(0)}}$ and $E_\mu^{(0)}$.  We expand $\ket{\mu_1}$ and $\ket{\mu_2}$ in $(|t_{r}|-1)$-th order perturbation theory. The missing order is contained in the coupling term $t_a \hat{V}_{k_\text{x}, k_\text{y}}$ (see Appendix~\ref{appendix_2}) in the matrix element $\braket{\mu_1|\hat{h}_{k_\text{x}, k_\text{y}}|\mu_2}$. Due to Eq.~(\ref{t_r_derivation}) we have $\mu_2 = \mu_1 - t_r \text{ mod } q$. Two cases have to be considered for its evaluation. The first case is $\mu_2 = \mu_1 + \text{sgn}(t_r)(q - |t_r|)$. Here we have to make use of the continuation condition $c_{\mu + q} = \e^{\ci k_\text{x} q} c_{\mu}$. This happens a total of $|t_r|$ times, namely whenever at a given degeneracy naively defining $\mu_2 = \mu_1 - t_r$ for $1\leq\mu_{1}\leq q$ would fail to provide $1\leq\mu_{2}\leq q$. Then
\begin{equation}\label{m_1_m_2_element}
\begin{aligned}[b]
\braket{\mu_1|\hat{h}_{k_\text{x}, k_\text{y}}|\mu_2} &\approx \frac{t_a^{|t_r|}}{(\Delta E)^{|t_{r}|-1}} \e^{- \text{sgn}(t_r) \ci k_\text{x} q} \\
&=: \varDelta_{r} \e^{-\ci \text{sgn}(t_r) k_\text{x} q}.
\end{aligned}
\end{equation}
$\Delta E$ is some constant $\propto t_b$ (see Eq.~(\ref{i_th_order_perturbation})). Note that it is not the gap between $E_{\mu_1}^{(0)}$ and $E_{\mu_2}^{(0)}$. Eq.~(\ref{effective_1}) then follows. In the other $q - |t_r|$ cases, where $\mu_2 = \mu_1 - t_r$ Eq.~(\ref{m_1_m_2_element}) yields $\varDelta_{r}$.

The reason that we get Dirac cones in Eq.~(\ref{dirac_sine}) and Eq.~(\ref{dirac_cosine}) is that $|t_r| = q/2$ and we get the same order in $t_a$ by raising $\ket{\mu_2^{(0)}}$ with $(\sum_\mu c_{\mu + 1}^{\dagger} c^{\pdagger}_{\mu})^{q/2}$ or lowering it with $(\sum_\mu c_{\mu + 1}^{\dagger} c^{\pdagger}_{\mu})^{-q/2}$. This results in two terms, where the continuation condition has to be used for one of the two. The $j$-th order of perturbation theory of a state $\ket{\mu}$ is
\begin{equation}\label{i_th_order_perturbation}
\begin{aligned}[b]
\ket{\mu^{(j)}} =& \displaystyle\sum_{\nu \neq \mu} \ket{\nu^{(0)}} \frac{\braket{\nu^{(0)}|\hat{V}|\mu^{(j-1)}}}{E_{\nu}^{(0)} - E_{\mu}^{(0)}} \\
&+ \displaystyle\sum_{l=1}^{j} E_{\mu}^{(l)} \displaystyle\sum_{\nu \neq \mu} \ket{\nu^{(0)}} \frac{\braket{\nu^{(0)}|\mu^{(j-l)}}}{E_{\nu}^{(0)} - E_{\mu}^{(0)}},
\end{aligned}
\end{equation}
with the orthogonality condition $\braket{\mu^{(j)}|\mu^{(0)}} = \delta_{j, 0}$ and $j$-th correction to the eigenenergies $E_{\mu}^{(j)} = -\braket{\mu^{(0)}|\hat{V}|\mu^{(j-1)}}$ for $j\geq 1$. Right at the degeneracy $E_{\mu}^{(0)} = -2t_b \cos(k_\text{y} + 2\pi \varphi \mu) = 0$, because the Dirac points have zero energy, so $E^{(0)}_{\mu+\nu} = -2t_b \cos(k_\text{y} + 2\pi \varphi \mu + 2\pi \varphi \nu) = + 2t_b \cos(k_\text{y} + 2\pi \varphi \mu - 2\pi \varphi \nu) = - E^{(0)}_{\mu-\nu}$. $\ket{\mu^{(j-1)}}$ in Eq.~(\ref{i_th_order_perturbation}) can be expressed as a weighted sum over states $\lbrace \ket{\nu^{(0)}} \rbrace$. By applying $\hat{V}$ to these states they will be lowered or raised with $\sum_{\lambda} c_{\lambda + 1}^{\dagger} c^{\pdagger}_{\lambda}$ and $\sum_{\lambda} c_{\lambda}^{\dagger} c^{\pdagger}_{\lambda + 1}$. Let $\ket{\nu_{\text{max}}^{(0)}}$ and $\ket{\nu_{\text{min}}^{(0)}}$ be the two states in the sum so that $|\mu - \nu_{\text{max/min}}| \geq |\mu - \nu|$. For every additional order in perturbation theory raising $\ket{\nu_{\text{max}}^{(0)}}$ will get an additional $(-1)$ compared to lowering $\ket{\nu_{\text{min}}^{(0)}}$. The gap scales proportional to $1/(\Delta E)^{q/2-1}$, so for $q = 4n$ we have to subtract the contributions from these raised and lowered states in the expansion of $\ket{\mu}$ at $E = 0$ and for $q = 2(2n+1)$ we have to add them. Then, we find for $q = 2(2n+1)$
\begin{equation}\label{hybridization_cos}
\braket{\mu_1|\hat{h}_{k_\text{x}, k_\text{y}}|\mu_2} \approx \frac{\varDelta}{2} (1 + \e^{\mp \ci k_\text{x} q}) = \varDelta \e^{\ci k_\text{x} q/2} \cos(k_\text{x} q/2)
\end{equation}
and for $q = 4n$
\begin{equation}\label{hybridization_sin}
\begin{aligned}[b]
\braket{\mu_1|\hat{h}_{k_\text{x}, k_\text{y}}|\mu_2} &\approx \frac{\varDelta}{2} (1 - \e^{\mp \ci k_\text{x} q}) \\
&= \pm \ci \varDelta \e^{\mp \ci k_\text{x} q/2} \sin(k_\text{x} q/2).
\end{aligned}
\end{equation}
We write $\pm$, because at $k_\text{y}$ and $k_\text{y} + \pi$ we find the same subspace $\ket{\mu_1}$ and $\ket{\mu_2}$ but a different sign in the exponent.

\section{Möbius bundle description of topological defects in 2D}\label{appendix_4}

Consider the toy Hamiltonian
\begin{equation}\label{appendix:dirac_cone}
\hat{H}^{+}(\vec{k}) = k_\text{x} \sigma_\text{x} + k_\text{y} \sigma_\text{y}.
\end{equation}
Around a circle $S^{1}$ at $\vec{k} = 0$ we may choose smooth complex eigenstates. These have a $U(1)$ gauge freedom, as we can multiply eigenstates with a complex phase factor $\e^{\ci \phi(\vec{k})}$. In mathematical terms eigenstates correspond to sections of a line bundle $E$~\cite{kau}. In the above case $E$ is given by the Cartesian product of the base space $S^{1}$ times the manifold associated to the structure group $U(1)$, the fiber, which is also $S^{1}$, because there exist eigenstates that are globally smooth. This results in the trivial bundle $E = S^{1} \times S^{1}$, which does not indicate by itself an associated topological invariant of this problem. Let us now consider the Hamiltonian
\begin{equation}\label{appendix:real_dirac_cone}
\hat{H}_{\mathcal{R}}^{+}(\vec{k}) = k_\text{x} \sigma_\text{x} + k_\text{y} \sigma_\text{z},
\end{equation}
which can be obtained from Eq.~(\ref{appendix:dirac_cone}) by unitary rotation. Since Eq.~(\ref{appendix:real_dirac_cone}) is a real Hermitian matrix, we are allowed to choose real eigenstates. However, it turns out that these cannot be made globally smooth. If we restrict ourselves to a $\mathbb{Z}_{2} \simeq \lbrace +1, -1 \rbrace$ structure group, then sections are defined on a Möbius bundle with fiber $[-1,1]$, which locally looks like $U_{j} \times [-1,1]$, where $\lbrace U_{j} \rbrace$ are open sets that cover $S^{1}$~\cite{kli, kir}. Normalized eigenstates are then sections, which are defined on the boundary of the Möbius band $\partial [-1, 1] = \mathbb{Z}_2$. This allows the definition of a $\mathbb{Z}_{2}$ invariant of a topological defect. Let $\ket{\Psi(\theta)}$ be an eigenstate of Eq.~(\ref{appendix:real_dirac_cone}), where $\theta \in S^{1}$ is a parametrization of the base space. With Eq.~(\ref{general_2_times_2}) one will find that real eigenstates obey $\ket{\Psi(0)} = -\ket{\Psi(2\pi)}$, so the bundle is twisted once. If a closed curve encircles $n$ cones of the form of Eq.~(\ref{appendix:real_dirac_cone}), the twist will be $n$~mod~$2$. Generally, twists of real vector bundles can be expressed by Stiefel–Whitney classes~\cite{ahn} that also play a role in the $\mathbb{Z}_2$ invariants of topological insulators~\cite{kau2} and in higher order topology~\cite{ahn2}. Here, the quantization of the Berry phase~\cite{kli, kir} is given by the first Stiefel–Whitney class. The quantization of the Berry phase will hold for any Hamiltonian with Dirac cones if a real Möbius bundle construction as discussed above is possible, which is for example the case for graphene, as its Hamiltonian can also be made purely real by a continuous unitary transformation.

The relation between the twist of the bundle and the presence of Dirac cones can be further clarified, by considering a special solid torus parametrization of the $p/q = 1/2$ Hofstadter Hamiltonian as an example. We already have two parameters $k_{\text{x}}$ and $k_{\text{y}}$ that describe the surface of the torus. We include a third parameter $r$ and substitute the $f_{\text{z}}$ component of the Hofstadter model, written as $h = \vec{f} \vec{\sigma}$, by $4-4r + 2 r \cos k_{\text{y}}$, where we have set $t_b = 1$. We obtain a solid torus geometry
\begin{equation}
\begin{aligned}
x &= (2 + r \cos k_{\text{y}}) \cos k_{\text{x}} \\
y &= (2 + r \cos k_{\text{y}}) \sin k_{\text{x}} \\
z &= r \sin k_{\text{y}},
\end{aligned}
\end{equation}
where e.g. $0 \leq r \leq 1$. Then the Hamiltonian will not be degenerate for isolated points but instead for all points of a closed curve that is parameterized by the conditions $k_{\text{x}} = \pi/2$ and $r = 4/(4-2\cos k_{\text{y}})$. The curve crosses the surface of the torus ($r = 1$), where the Dirac points appear in the Brillouin zone of the $p/q=1/2$ Hofstadter model. Going around a single one of the Dirac cones results in a path around the curve of degeneracies, which is not homeomorphic to a path that does not encircle a Dirac cone. The difference between both scenarios can be captured by a $\mathbb{Z}_2$ invariant~\cite{berry2} which describes the topological charge of a nodal line~\cite{nodal}. The Möbius bundle of the Dirac cones of the Hofstadter model also corresponds to a real bundle as can be seen, when applying the unitary transformation
\begin{equation}
U = \begin{pmatrix}
1 & 0 \\
0 & e^{-ik_{\text{x}}}
\end{pmatrix}
\end{equation}
which brings the Hamiltonian to a purely real form. It gives the wave function a twisted boundary condition $\Psi(k_{\text{x}} = -\pi/2) = -\Psi(k_{\text{x}} = \pi/2)$ but otherwise leaves its geometric properties invariant, as $\diff (U^{\dagger} \diff U) = 0$.

\begin{figure}[b]
  \centering
  \includegraphics[width=1.0\linewidth]{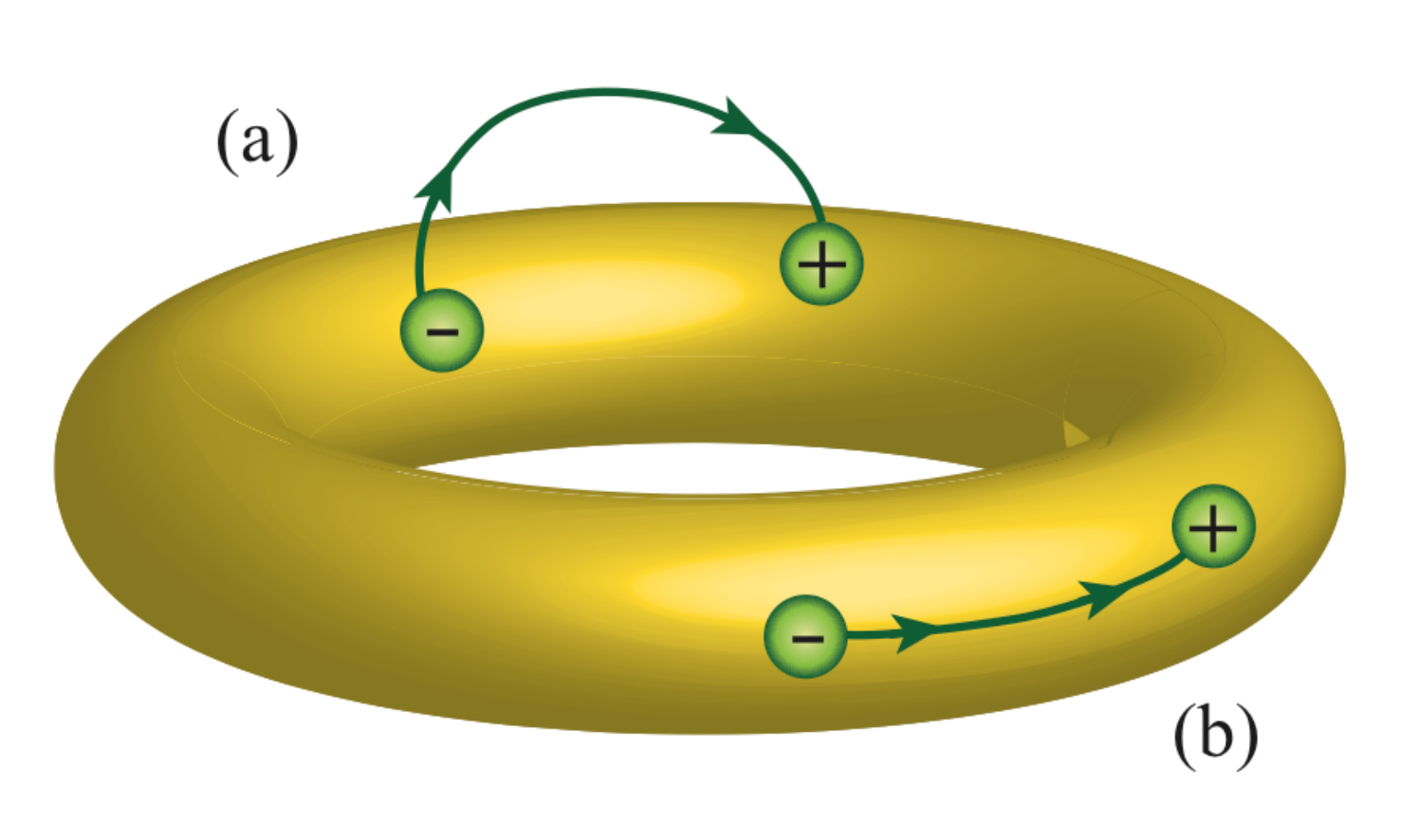}
  \caption{Flux tubes between monopoles in a torus geometry. (a) Flux tubes touching the torus surface only at the monopole positions which will lead to well-defined winding numbers. (b) Flux tubes lying exactly on the surface of the torus that will lead to a discontinuity of the eigenstates at the respective $\vec{k}$ values.}\label{fig:torus_moebius}
\end{figure}

A good graphical indicator of the twist of the Möbius bundle is the line between the Dirac cones in Fig.~\ref{subfig:q_2_top} and \ref{subfig:q_2_sub}, where the vortex field is discontinuous. This is due to the condition $\ket{\Psi(0)} = -\ket{\Psi(2\pi)}$ in the chosen gauge and the fact that the winding vectors of the Dirac cones do not have a $\sigma_{\text{z}}$ component. In this particular situation and with a solid torus parametrization, where the two Dirac points appear as isolated monopoles, they will have opposite charge (the parametrization given above for the Hofstadter model would not be suitable here, as it leads to an equivalent of a real vector bundle. In this case, the Chern classes which are commonly used to express the charge of Weyl points, are all necessarily trivial and the solid torus parametrization could not lead to isolated degeneracies with linear dispersion). They are connected by a flux tube, which lies exactly on the surface of the torus and which causes the discontinuity in the states, see Fig.~\ref{fig:torus_moebius}. An infinitesimal rotation of the Hamiltonian and the same gauge convention of the eigenstates will push the flux tube away from the torus surface everywhere, except at the monopole positions, which will correspond to an opposite, well-defined winding of the two Dirac cones.

It should also be noted that this Möbius bundle construction is possible for any (up to a gauge transformation) real Hamiltonian on a manifold with nontrivial first homology if some non-contractible loop carries a Berry phase of $\pi~(\text{mod}~2\pi)$. In the vortex field this manifests itself in a discontinuity extending along a non-contractible 1D line. When this discontinuity in the vortex field is crossed, the corresponding eigenstate obtains a phase shift of $\pi$~\cite{rad}.

\bibliography{ms}

\end{document}